\documentclass[a4paper,10pt]{article}

\usepackage{lmodern}
\usepackage[T1]{fontenc}
\usepackage[utf8]{inputenc}
\usepackage{amsmath}
\usepackage{booktabs}
\usepackage{graphicx}
\usepackage{xcolor}
\usepackage[textwidth=153mm,textheight=255mm,centering]{geometry}
\usepackage[font=small,labelfont=bf,labelsep=period]{caption}
\usepackage[colorlinks=true,allcolors=blue]{hyperref}
\usepackage{microtype}

\newcommand{\etal}{\textit{et al}}

\let\oldtabular\tabular
\renewcommand{\tabular}{\fontsize{8}{10}\selectfont\oldtabular}

\newcommand{\articletype}[1]{%
  \begin{center}\small\scshape #1\end{center}\vspace{2mm}}

\renewcommand{\title}[1]{%
  {\exhyphenpenalty=10000\hyphenpenalty=10000
   \begin{center}\LARGE\bfseries #1\par\end{center}}\vspace{3mm}}

\renewcommand{\author}[1]{%
  \begin{center}\large #1\par\end{center}\vspace{1mm}}

\newcommand{\affil}[1]{%
  \begin{center}\small #1\par\end{center}}

\newcommand{\email}[1]{%
  \begin{center}\small E-mail: \href{mailto:#1}{#1}\par\end{center}\vspace{4mm}}

\newcommand{\keywords}[1]{%
  \vspace{2mm}\noindent{\small\textbf{Keywords:} #1}\par\vspace{2mm}}

\newcommand{\orcid}[1]{%
  \,{\small(ORCID \href{https://orcid.org/#1}{#1})}}

\newcommand{\ack}[1]{\section*{Acknowledgments}#1}
\newcommand{\funding}[1]{\section*{Funding}#1}
\newcommand{\roles}[1]{\section*{Author contributions}#1}
\newcommand{\data}[1]{\section*{Data availability}#1}

\begin{document}

\articletype{Topical Review}

\title{Fractional-order hardware for neuromorphic computing: Is the
order really the problem?}

\author{Christof Teuscher\orcid{0000-0003-3853-8018}}

\affil{Department of Electrical and Computer Engineering, Portland
State University, Portland, Oregon, United States of America}

\email{teuscher@pdx.edu}


\begin{abstract}
Does a neuromorphic system need a true power-law memory kernel, and if
so, can anyone build one? Neuromorphic systems are increasingly asked to
process signals whose structure spans many timescales at once, from the
millisecond dynamics of a spike to the tens of seconds over which a
sensory neuron adapts. Integer-order circuits buy each additional
timescale with an additional state variable. Fractional-order dynamics
offer a different bargain: one operator whose power-law kernel carries a
continuum of timescales, tuned by a single continuous parameter, the
order $\alpha$. What makes that attractive also makes it expensive. A
fractional derivative is non-local, so evaluating it faithfully costs
storage and arithmetic that grow with the retained history, where an
integer-order derivative costs a constant.

In this review we organize the hardware literature around that cost. We
derive the retained history required to hold the truncation error below
a tolerance $\varepsilon$, show that it scales as
$\varepsilon^{-1/\alpha}$, and set beside it a second and independent
limit on the direct form: in fixed point the weights themselves
underflow, so word length caps the usable history no matter how long the
buffer is. The two limits move at very different rates with the order,
and where they cross is what decides whether a word length can serve an
order at all. We then use both results to sort published hardware into
three strategies, note a fourth that the numerical literature has
developed and this hardware has not adopted, and survey digital, analog
and device-level work against them. Along the way we ask whether the field is worried about
the right obstacle. It is not. Fabricated constant-phase devices already
span the orders that two groups, sweeping the order on the same
benchmark, identify as task-optimal. The
order gap that the literature frets over has largely closed, with a
residual gap for long-memory tasks whose optimum sits near $0.1$ and for
the lower order that describes cortical adaptation. What
remains is a frequency-band gap of roughly three decades at the low end,
where the slowest neuromorphic adaptation lives. That corner is not
empty, since electrochemical double-layer electrodes work there, but
every device in it is a discrete component that no neuromorphic process
can integrate, and no integrable thin-film element has been characterized
there at all.
\end{abstract}

\section{Introduction}
\label{sec:intro}

A neuron adapts to a step change in its input over milliseconds, over
seconds, and over tens of seconds, all at the same time. Recordings
from cortical, auditory and visual neurons show adaptation that decays
as a power law rather than as a single exponential, which is to say
that no one time constant describes it \cite{Thorson74, Fairhall01,
Ulanovsky04, Wark07}. Engineering a system with the same property is
awkward. In an integer-order model, every additional timescale costs an
additional state variable and an additional set of parameters to fit,
so a circuit that adapts over four decades of time carries the machinery
for four decades of time explicitly.

Fractional calculus offers a different arrangement. The fractional
derivative $D^{\alpha}$ of non-integer order $\alpha$ acts on a signal
through a power-law memory kernel, and a single such operator produces
relaxation spread continuously over timescales rather than concentrated
at one. The order $\alpha$ is a continuous parameter with no
integer-order counterpart: it does not select between behaviors from a
discrete menu, it interpolates. For a designer accustomed to buying
temporal richness one state variable at a time, that is an appealing
proposition, and it is the reason fractional-order models keep
reappearing in neuronal modeling \cite{Lundstrom08, Teka14, Teka16,
Weinberg15} and in circuit theory \cite{Westerlund94, Elwakil10,
RadwanSalama12}.

The proposition comes with a bill. The same non-locality that spreads
relaxation across timescales means that evaluating $D^{\alpha}x(t)$
requires the trajectory $x$ over its whole past, not merely its present
value. Implemented directly, this costs storage and arithmetic
proportional to the number of retained samples, per state variable, per
timestep, where an integer-order derivative costs a constant. Every
piece of fractional-order hardware in the literature is, at bottom, a
decision about how to pay that bill, whether by truncating the history,
by replacing the operator with a finite integer-order approximation, or
by finding a physical process whose native relaxation is already
power-law. We take that decision as the organizing principle of this
review.

So why has nobody built one? Part of the answer is that the work
relevant to fractional-order neuromorphic hardware is scattered across
communities that do not read each other. Numerical
analysts have characterized the error incurred by truncating the memory
of a fractional operator \cite{Deng07, Xu11, Hai22}. Circuit designers
have built digital and analog approximations and reported their
resource cost \cite{Monir22, Tolba19a, Charef06, Tsirimokou17}.
Electrochemists and materials scientists have fabricated devices whose
impedance holds a constant phase over decades of frequency
\cite{Elshurafa13, Agambayev18, John17}. Computational neuroscientists
have shown that neurons perform something close to fractional
differentiation and have built neuron models on that basis
\cite{Lundstrom08, Teka14}. Each of these communities uses different vocabulary for the same
quantity, and they rarely cite one another. The question a hardware
designer actually faces, which is what a fractional operator costs and
what a physical device can deliver, has not been posed in one place.

This review sets out to pose it. We ground the case in the experimental
neuroscience of power-law adaptation, derive and quantify the cost of
non-locality rather than asserting it, and ask what the published device
data can and cannot currently support. On the last point the answer
turns out to be uncomfortable, and section~\ref{sec:mismatch} is where
we say so.

Some delimitation is necessary. Fractional-order control is a large and
mature field with its own reviews \cite{Ali24, Podlubny02}, and we
treat it only where a result bears on neuromorphic implementation.
Fractional-order chaotic systems studied for their dynamics alone, and
their application to image encryption, are outside our scope, as is the
mathematical theory of fractional differential equations beyond what a
hardware designer needs. We do not restrict the order to $\alpha\in(0,1)$. That range covers most
of the literature and every fabricated device, but the interval $(1,2)$
turns out to invert the review's central trade rather than merely extend
it, and section~\ref{sec:models} says why.

One further exclusion deserves more than a clause, because it is the
oldest and largest application of fractional calculus and a reader may
reasonably expect it here. Linear viscoelasticity has used fractional
operators since the middle of the last century
\cite{Koeller84, Mainardi10}. Its basic element, a constitutive law in
which stress is proportional to a fractional derivative of strain,
interpolates continuously between a spring and a dashpot exactly as
$\alpha$ interpolates between a resistor and a capacitor here, and the
resulting fractional Maxwell and Zener models are the mechanical twins
of the Cole--Cole forms used for membrane impedance. The
correspondence is not an analogy but the same mathematics: in both
domains a power-law kernel is what a broad distribution of relaxation
times produces, whether those times come from polymer chain
reorientation or from interfacial charge transport. The connection
reaches biology as well. Living cells are themselves power-law
viscoelastic, with an exponent near $0.17$ that is common to five very
different cell types and holds from $0.1$ to $100$\,Hz
\cite{Fabry01}, which is close to the order reported for cortical
fractional differentiation \cite{Lundstrom08} and lies in the frequency
band that, as section~\ref{sec:mismatch} shows, no fabricated
fractional-order device presently covers.

We nonetheless leave that literature out. Its concerns are mechanical
rather than electrical, none of it targets hardware, and treating it
adequately would double the length of this review. Two things are worth
flagging for anyone who does take it up. The resemblance between the
cell-mechanical exponent and the neuronal one is most likely a common
consequence of broad relaxation spectra in disordered media rather than
evidence of a shared mechanism, and it should not be read as one
without direct evidence. More substantively, the numerical methods
developed for fractional viscoelasticity are a genuinely distinct
family from the frequency-domain rational approximations reviewed in
section~\ref{sec:strategies}. Diffusive representations recast the
fractional operator as a bank of ordinary first-order relaxations with
fixed weights \cite{Diethelm22, ChaudharyDiethelm23}, which is a
time-domain substitution rather than a frequency-domain one, and which
maps onto a substrate already built from leaky integrators.
Section~\ref{sec:strategies} takes that family up as a fourth strategy
rather than leaving it here, because it turns out to bound the cost
argument on which this review is built.

\subsection{Scope, and how this corpus was assembled}
\label{sec:method}

This is a narrative review and not a systematic one, and the distinction
matters enough to state. The corpus began from an existing classification
of the fractional-order circuits and control literature, was extended by
citation chasing forward and backward from that seed and from the
neuroscience and device-physics work it pointed to, and was closed when
new searches stopped returning implementations the tables did not already
contain. We did not run a protocol-registered database search, and we
make no claim to have enumerated the field.

One family is deliberately cited but not tabulated. The electrochemical
double-layer literature is large, it predates this field, and its
impedance spectroscopy is performed to different conventions;
section~\ref{sec:edlc} says what follows from leaving it out.

This review is addressed to two audiences that currently overlap very
little: neuromorphic engineers who are unfamiliar with fractional
calculus, and researchers in fractional-order circuits who are
unfamiliar with the constraints of neuromorphic hardware. We therefore
begin in section~\ref{sec:operators} with a self-contained treatment of
the fractional operators, the memory kernel, and the cost of evaluating
them, assuming no prior exposure. Section~\ref{sec:biology} reviews the
experimental evidence that biological neurons perform fractional
differentiation, and the counterargument that a small number of
exponential processes may account for the same observations.
Section~\ref{sec:models} surveys fractional-order neuron and network
models. Sections~\ref{sec:digital} and~\ref{sec:devices} review digital
and analog implementations and devices with intrinsic fractional
dynamics, respectively. Section~\ref{sec:gaps} identifies what the
field does not currently measure, and section~\ref{sec:outlook} sets
out research targets.

\section{Fractional operators and what they cost}
\label{sec:operators}

This section is written for a reader with no background in fractional
calculus. We introduce only the material the rest of the review uses:
the three standard definitions and the reason a hardware designer
should care which one is chosen, the memory kernel and its power-law
tail, and the storage and arithmetic cost of evaluating the operator.
Readers already familiar with the formalism may wish to begin at
section~\ref{sec:cost}, which contains the cost analysis that organizes
the remainder of the review.

\subsection{Definitions, and which one you actually build}

Fractional calculus generalizes differentiation and integration to
non-integer order. There is no single generalization; there is a family
of them, agreeing under conditions that are usually but not always met
in practice. Three definitions dominate the engineering literature
\cite{Podlubny99, Petras11}.

The Riemann--Liouville derivative of order $\alpha\in(0,1)$ with lower
terminal $a$ is defined by differentiating a fractional integral,
\begin{equation}
  {}^{\mathrm{RL}}_{a}D^{\alpha}_{t}x(t)
  = \frac{1}{\Gamma(1-\alpha)}\,\frac{\mathrm{d}}{\mathrm{d}t}
    \int_{a}^{t}\frac{x(\tau)}{(t-\tau)^{\alpha}}\,\mathrm{d}\tau .
  \label{eq:rl}
\end{equation}
The Caputo derivative reverses the order of the two operations,
differentiating first and integrating afterwards,
\begin{equation}
  {}^{\mathrm{C}}_{a}D^{\alpha}_{t}x(t)
  = \frac{1}{\Gamma(1-\alpha)}
    \int_{a}^{t}\frac{\dot{x}(\tau)}{(t-\tau)^{\alpha}}\,\mathrm{d}\tau .
  \label{eq:caputo}
\end{equation}
The difference is not cosmetic. An initial-value problem posed with the
Riemann--Liouville derivative requires initial conditions expressed as
fractional derivatives of $x$, which have no direct physical reading;
the Caputo form takes ordinary initial conditions such as a membrane
voltage at $t=0$. For that reason essentially all fractional neuron
models are posed in the Caputo sense, and we adopt it throughout when
writing dynamics.

The Grünwald--Letnikov derivative takes a different route, generalizing
the backward difference quotient rather than an integral:
\begin{equation}
  {}^{\mathrm{GL}}_{a}D^{\alpha}_{t}x(t)
  = \lim_{h\to0}h^{-\alpha}\sum_{j=0}^{\lfloor (t-a)/h\rfloor}
    (-1)^{j}\binom{\alpha}{j}\,x(t-jh).
  \label{eq:glcont}
\end{equation}
For functions that are sufficiently smooth on $[a,t]$, the
Grünwald--Letnikov and Riemann--Liouville derivatives coincide, and the
Caputo derivative differs from both by terms involving the initial
conditions, which vanish for $x(a)=0$ \cite{Podlubny99}. The practical
consequence is a convenient division of labour: dynamics are posed in
the Caputo sense because the initial conditions are physical, and
evaluated in the Grünwald--Letnikov sense because
equation~(\ref{eq:glcont}) is already a discrete sum and therefore
directly implementable. Almost every digital realization reviewed in
section~\ref{sec:digital} does exactly this.

There is a second and less obvious reason to evaluate in the
Gr\"unwald--Letnikov sense once the history has been truncated. Olivares
and Santamaria report that a truncated Caputo approximation in a spiking
model is sensitive to the discontinuity at each spike and can produce
spurious spontaneous firing, where the Gr\"unwald--Letnikov form, being
discrete from the outset, is not \cite{Olivares26}. For a neuron model the
choice of definition is therefore not only about initial conditions.

Two cautions are worth stating now, because they surprise designers
coming from integer-order practice. First, fractional derivatives do
not in general compose: $D^{\alpha}D^{\beta}\ne D^{\alpha+\beta}$
unless additional conditions hold, so a half-order differentiator
applied twice is not a first-order differentiator. Second, the operator
depends on the lower terminal $a$, which is to say on when the system
is deemed to have started. There is no fractional analog of resetting
a state variable, a point we return to in section~\ref{sec:devices}
when the memory lives in a physical relaxation process that cannot be
cleared.

\subsection{The memory kernel}

Both equations~(\ref{eq:rl}) and~(\ref{eq:caputo}) weight the past by
$(t-\tau)^{-\alpha}$. This kernel is the whole story. It decays, so
recent history matters more than remote history, but it decays as a
power law, so remote history never becomes negligible in the way it
does under an exponential kernel. A first-order low-pass filter forgets
its input exponentially and can therefore be summarized by a single
state; a fractional operator cannot.

\begin{figure}
  \centering
  \includegraphics[width=\textwidth]{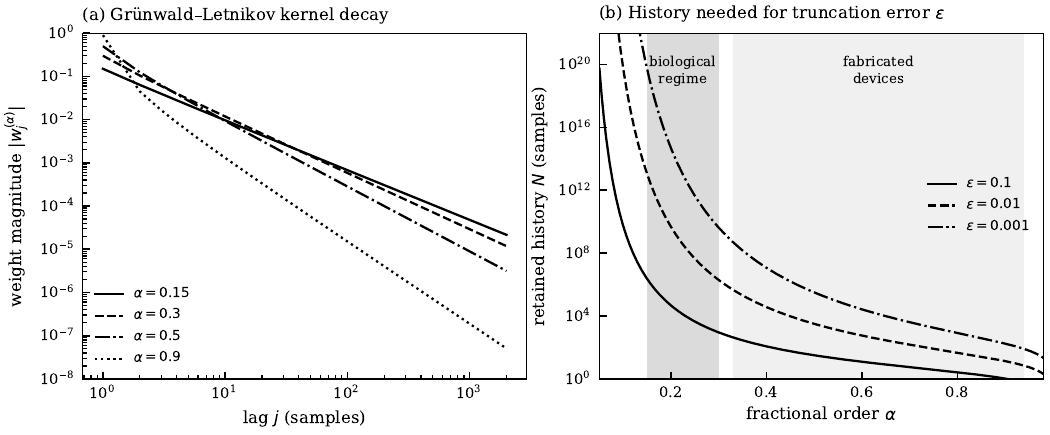}
  \caption{(a) Magnitude of the Gr\"unwald--Letnikov weights
  $|w_j^{(\alpha)}|$ of equation~(\ref{eq:recurrence}) against lag $j$,
  showing power-law rather than exponential decay. Smaller $\alpha$
  gives a heavier tail. (b) Retained history $N$ required to hold the
  truncation error below $\varepsilon$, from equation~(\ref{eq:Neps}).
  Shading marks the fractional orders reported for cortical neurons and
  required by models of biological adaptation
  ($\alpha\approx0.15$--$0.3$, section~\ref{sec:biology}) and the full
  range spanned by the fabricated constant-phase devices of
  table~\ref{tab:devices} ($\alpha=0.33$--$0.94$,
  section~\ref{sec:devices}). The digitally affordable regime does not
  overlap the range that describes cortical adaptation, but it overlaps
  most of what devices deliver. Whether the remaining gap binds a given
  design depends on which order that design is targeting, which
  section~\ref{sec:mismatch} takes up.}
  \label{fig:kernel}
\end{figure}

The order $\alpha$ controls how heavily the tail is weighted. As
$\alpha\to1$ the kernel concentrates near $\tau=t$ and the operator
approaches an ordinary derivative. As $\alpha\to0$ the weighting
flattens and the operator approaches the identity, retaining
essentially all of its history. Small $\alpha$ therefore means long
memory, and, as the next section makes quantitative, expensive memory.
Figure~\ref{fig:kernel}(a) shows this directly in the discrete weights.

\subsection{The cost of non-locality}
\label{sec:cost}

Truncating equation~(\ref{eq:glcont}) at $N$ retained samples with step
$h$ gives the form that is actually implemented,
\begin{equation}
  D^{\alpha}x(t)\;\simeq\;h^{-\alpha}\sum_{j=0}^{N}w_j^{(\alpha)}\,
  x(t-jh),
  \qquad
  w_j^{(\alpha)}=(-1)^j\binom{\alpha}{j},
  \label{eq:gl}
\end{equation}
with the weights generated by the recurrence
\begin{equation}
  w_0^{(\alpha)}=1,
  \qquad
  w_j^{(\alpha)}=w_{j-1}^{(\alpha)}\left(1-\frac{1+\alpha}{j}\right).
  \label{eq:recurrence}
\end{equation}

Equation~(\ref{eq:recurrence}) matters practically. The weights cost
$\mathcal{O}(N)$ to compute once and can then be stored and shared
across every state variable that uses the same order. The history
cannot be shared: each state variable needs its own buffer of $N$ past
samples and its own $N$ multiply--accumulate operations at every
timestep. Evaluating equation~(\ref{eq:gl}) therefore costs
$\mathcal{O}(N)$ storage and $\mathcal{O}(N)$ arithmetic per state
variable per timestep, against $\mathcal{O}(1)$ for an integer-order
derivative. That asymmetry between shared coefficients and unshared
history is the reason fractional operators scale badly with network
size, and it is what the strategies in section~\ref{sec:strategies}
respond to. All of that describes the direct form, which is what every
implementation surveyed here builds.
Section~\ref{sec:strategies} takes up schemes that compute the same
operator in $\mathcal{O}(\log N)$ storage, and it is those, not
$\mathcal{O}(N)$, that bound what a device can save.

The obvious question is how large $N$ must be. The weights decay as a
power law,
\begin{equation}
  \bigl|w_j^{(\alpha)}\bigr|\;\sim\;
  \frac{j^{-(1+\alpha)}}{\bigl|\Gamma(-\alpha)\bigr|},
  \qquad j\to\infty,
  \label{eq:decay}
\end{equation}
so truncating to $N$ terms discards a tail of total weight
\begin{equation}
  T(N)=\sum_{j>N}\bigl|w_j^{(\alpha)}\bigr|
  \;\simeq\;\frac{N^{-\alpha}}{\alpha\bigl|\Gamma(-\alpha)\bigr|}
  \;=\;\frac{N^{-\alpha}}{\bigl|\Gamma(1-\alpha)\bigr|},
  \label{eq:tail}
\end{equation}
where the second equality uses $\alpha\,\Gamma(-\alpha)=-\Gamma(1-\alpha)$.
Because $\sum_{j\ge1}|w_j^{(\alpha)}|=1$ exactly, $T(N)$ is the
fraction of the total weight discarded, so for a bounded signal
$|x|\le M$ the error in the weighted sum is at most $M\,T(N)$. The error
in $D^{\alpha}x$ itself is larger by the prefactor that
equation~(\ref{eq:gl}) carries outside the sum, at most
$h^{-\alpha}M\,T(N)$. We read $\varepsilon$ throughout as a relative
tolerance on the weighted sum, which is what makes
equation~(\ref{eq:Neps}) independent of $h$; converting it to an
absolute error on the derivative costs that prefactor, which is about
$32$ at $h=1$\,ms and $\alpha=0.5$ and about $500$ at $\alpha=0.9$. Equation~(\ref{eq:tail})
is the short-memory bound of Podlubny \cite{Podlubny99} arrived at from
the weights rather than from the integral, and the two agree. We
verified the asymptotic form against exact partial sums at $N=10^{5}$
and found agreement to five significant figures for $\alpha=0.15$,
$0.5$ and $0.9$.

Inverting equation~(\ref{eq:tail}) gives the retained history required
to meet a truncation error $\varepsilon$,
\begin{equation}
  N(\alpha,\varepsilon)=
  \left(\frac{1}{\varepsilon\bigl|\Gamma(1-\alpha)\bigr|}\right)^{1/\alpha}.
  \label{eq:Neps}
\end{equation}

The exponent is $1/\alpha$. This is the result on which the rest of the
review turns. The cost of accuracy is not merely larger at small
$\alpha$; it changes character. Table~\ref{tab:cost} evaluates
equation~(\ref{eq:Neps}). At $\alpha=0.9$, one percent truncation error
needs a fourteen-sample window, which is a trivial shift register. At
$\alpha=0.5$ it needs about 3200 samples, which is a design decision. At
$\alpha=0.3$ it needs of order $10^{6}$, and at $\alpha=0.15$ the bound
asks for $10^{13}$, which is a way of saying that truncation is not a
viable strategy in that regime at any sample rate one would build.
Figure~\ref{fig:kernel}(b) shows the same result across the range.

Note the direction of the exponent before reading the table: the cost
falls as $\alpha$ rises, and section~\ref{sec:models} follows that
observation past $\alpha=1$, where it leads somewhere the device
literature cannot go. Two qualifications belong with this. The bound is worst-case over
bounded signals, so a particular signal will usually do better than
table~\ref{tab:cost} indicates. And $T(N)$ measures truncation of the
kernel, not end-to-end error of a model being integrated, which also
depends on the dynamics. Neither qualification touches the scaling: the
exponent $1/\alpha$ is a property of the operator and not of the signal
or the system.

The consequence for this review is stated here and returned to in
section~\ref{sec:mismatch}. Section~\ref{sec:biology} will show that the
fractional orders reported for cortical neurons, and required by models
that reproduce biological adaptation, lie in the range
$\alpha\approx0.15$ to $0.3$. That is precisely the range in which
equation~(\ref{eq:Neps}) says digital truncation stops being affordable.

It matters, however, which order a design is actually obliged to hit.
The biological order is a descriptive fact about cortical neurons. It is
not the same quantity as the order that maximizes performance on a
given workload, and the two need not coincide. Sweeping $\alpha$ across
three benchmarks, Mastin and colleagues report an optimum near
$\alpha=0.3$--$0.5$ for spoken-digit classification, $0.5$--$0.7$ for
cart-pole control, and near $\alpha=0.1$ for a physiological prediction
task whose performance falls monotonically with $\alpha$. Two of those
three optima sit in the range where truncation remains affordable and
where fabricated devices already operate. The third does not. The cost
result of this section therefore does not condemn the digital route
outright; it says that the route is affordable for task-optimal orders
on some workloads and unaffordable for the long-memory tasks, and it
gives a quantitative criterion for telling the two apart in advance.

The worst-case character of the bound also has to be kept in view.
Equation~(\ref{eq:Neps}) asks for $N\approx3200$ at $\alpha=0.5$ and
$\varepsilon=10^{-2}$. Empirically, on the spoken-digit task at the same
order, classification accuracy saturates by a retained history of about
$150$ samples, with $L=100$ already at the knee of the curve
\cite{Mastin26}. The gap of more than an order of magnitude is not a
contradiction. The bound covers every bounded signal, while a particular
task only needs the kernel resolved over the timescales its own inputs
occupy. Read equation~(\ref{eq:Neps}) as a ceiling to measure against, not as a
specification.

Olivares and Santamaria bracket the same trade from the other end. They
report their fractional networks consuming about twenty times the
arithmetic of an integer-order baseline per second of control, and report
it as flat across every order tested, because the cost is set by the fixed
length of the retained history rather than by $\alpha$ \cite{Olivares26}.
Once the window is chosen, the order stops mattering to the bill.

That said, a task metric can hide what the model itself is doing, and
Teka and colleagues supply the cautionary case. Running a fractional
integrate-and-fire neuron at $\alpha=0.2$, they removed the accumulated
history at a chosen time and measured the effect on firing rate. Even
when the removal was deferred to $100$\,s, the change in rate stayed
large \cite{Teka17}. Classification accuracy on a spoken digit is
forgiving of a truncated tail in a way that the neuron's own spike timing
is not. Which of the two a designer should care about depends on whether
the hardware is meant to classify or to reproduce a dynamical
behavior, and the two answers point to retained histories that differ by
more than an order of magnitude.

\begin{table}
\caption{Retained history $N$ required by equation~(\ref{eq:Neps}) to
hold the Gr\"unwald--Letnikov truncation error below $\varepsilon$,
for a bounded signal. The asymptotic form of equation~(\ref{eq:tail})
was verified against exact partial sums to five significant figures at
$N=10^{5}$.}
\centering
\begin{tabular}{lcccc}
\hline
$\alpha$ & $\Gamma(1-\alpha)$ & $\varepsilon=10^{-1}$ & $\varepsilon=10^{-2}$ & $\varepsilon=10^{-3}$ \\
\hline
0.15 &  1.1125 & $2.28\times10^{6}$  & $1.06\times10^{13}$ & $4.91\times10^{19}$ \\
0.20 &  1.1642 & $4.68\times10^{4}$  & $4.68\times10^{9}$  & $4.68\times10^{14}$ \\
0.30 &  1.2981 & $9.0\times10^{2}$   & $1.95\times10^{6}$  & $4.19\times10^{9}$  \\
0.50 &  1.7725 & $32$                & $3.18\times10^{3}$  & $3.18\times10^{5}$  \\
0.70 &  2.9916 & $5.6$               & $150$               & $4.04\times10^{3}$  \\
0.82 &  5.1318 & $2.3$               & $37$                & $620$               \\
0.90 &  9.5135 & $1.1$               & $14$                & $176$               \\
0.95 & 19.470  & $<1$                & $5.6$               & $63$                \\
\hline
\end{tabular}
\label{tab:cost}
\end{table}

\subsection{Coefficient precision, a second and independent limit}
\label{sec:precision}

Equation~(\ref{eq:Neps}) bounds the history a design must \emph{retain}.
It says nothing about whether that history can be \emph{represented}.
In a fixed-point implementation the weights themselves are stored to
finite precision, and because $|w_j^{(\alpha)}|$ decays as
$j^{-(1+\alpha)}$ there is an index beyond which every weight rounds to
zero. Past that point a longer buffer buys nothing at all: the samples
are still there, but they are multiplied by zero.

Write $\mathrm{UQ}0.b$ for an unsigned fixed-point format carrying $b$
fractional bits and no integer bits, and let $k_{\max}(\alpha,b)$ be the
largest index whose weight remains representable, that is
$|w_k^{(\alpha)}|\geq2^{-b}$. A truncated Gr\"unwald--Letnikov operator
is then useful only where
\begin{equation}
k_{\max}(\alpha,b)\;\geq\;N(\alpha,\varepsilon),
\label{eq:feasible}
\end{equation}
and the two sides behave very differently in $\alpha$. Both fall as the
order rises, but $N$ falls by nineteen orders of magnitude across
$\alpha\in[0.1,0.9]$ while $k_{\max}$ falls by less than two. The curves
therefore cross, and the crossing is what determines whether a given
word length can serve a given order at all.

Table~\ref{tab:precision} evaluates both sides at
$\varepsilon=10^{-2}$. The representable-weight index is not itself new,
and our values agree with the fixed-point table of \cite{Mastin26} as
the caption records. What the literature does not state is where that
index crosses the history the tolerance demands, which is what the last
column reports. Eight-bit coefficients serve nothing below
$\alpha=0.97$. That is to say they serve nothing. Sixteen bits reach down to
$\alpha=0.62$, twenty-four bits to $0.38$, and thirty-two bits to
$0.27$. Below roughly $\alpha=0.27$ no practical word length suffices,
because the required history outruns the representable history however
many bits are spent. That holds for the direct form.
Section~\ref{sec:strategies} describes schemes that aggregate the far
tail before storing it, so the individual weights never underflow and
the rule does not carry over to them.

Two consequences follow for the hardware surveyed in
section~\ref{sec:digital}. The published designs sit squarely in the
constrained region: \cite{Tolba19a} and \cite{Malik20} carry
twenty-four fractional bits, \cite{Rana16} seventeen, and the
fractional integrate-and-fire neuron of \cite{Mastin26} sixteen. At one
percent truncation error those formats reach $\alpha\geq0.38$,
$\alpha\geq0.58$ and $\alpha\geq0.62$ respectively, so none of them
could reach the biological order even if given unlimited memory. And
the order that section~\ref{sec:biology} reports for cortical neurons,
near $\alpha=0.15$, is now excluded twice over and for independent
reasons: the retained history it demands is unbuildable, and the
coefficients it demands are unrepresentable. A designer who solves the
memory problem, by whatever means, still meets the second wall.

The caveat of section~\ref{sec:cost} carries over unchanged. Both sides
of equation~(\ref{eq:feasible}) are worst-case, and a particular task
will usually do better than the table says. What does not change is
that the two limits are independent, so satisfying one gives no
purchase on the other.

\begin{table}
\caption{Retained history required by equation~(\ref{eq:Neps}) at
$\varepsilon=10^{-2}$, against the largest representable weight index
$k_{\max}$ in three fixed-point formats. Values of $k_{\max}$ are from
the exact weight recurrence and agree with the fixed-point table of
Mastin \etal{} \cite{Mastin26} in nine of its ten entries; the tenth,
$\alpha=0.1$ at $\mathrm{UQ}0.32$, is printed there as $4.18\times10^{7}$
where three independent computations give $6.64\times10^{7}$. That row is infeasible at every word length either way, so
nothing here turns on the difference. The last column
gives the shortest word length satisfying
equation~(\ref{eq:feasible}). Both sides describe the direct form;
section~\ref{sec:strategies} gives the schemes to which the rule does
not apply.}
\centering\footnotesize
\begin{tabular}{lrrrrl}
\hline
$\alpha$ & $N(\alpha,10^{-2})$ & $k_{\max}$, $\mathrm{UQ}0.16$ & $k_{\max}$, $\mathrm{UQ}0.24$ & $k_{\max}$, $\mathrm{UQ}0.32$ & shortest usable \\
\hline
0.1 & $5.2\times10^{19}$ & 2\,775 & 429\,170 & $6.6\times10^{7}$ & none \\
0.2 & $4.7\times10^{9}$  & 2\,378 & 241\,597 & $2.5\times10^{7}$ & none \\
0.3 & $2.0\times10^{6}$  & 1\,643 & 116\,980 & $8.3\times10^{6}$ & $\mathrm{UQ}0.32$ \\
0.4 & $3.7\times10^{4}$  & 1\,078 & 56\,589  & $3.0\times10^{6}$ & $\mathrm{UQ}0.24$ \\
0.5 & $3.2\times10^{3}$  & 699    & 28\,189  & $1.1\times10^{6}$ & $\mathrm{UQ}0.24$ \\
0.6 & 571                & 452    & 14\,472  & $4.6\times10^{5}$ & $\mathrm{UQ}0.24$ \\
0.7 & 150                & 290    & 7\,564   & $2.0\times10^{5}$ & $\mathrm{UQ}0.16$ \\
0.8 & 47                 & 179    & 3\,910   & $8.5\times10^{4}$ & $\mathrm{UQ}0.16$ \\
0.9 & 14                 & 99     & 1\,834   & $3.4\times10^{4}$ & $\mathrm{UQ}0.16$ \\
\hline
\end{tabular}
\label{tab:precision}
\end{table}

\subsection{Three ways to pay}
\label{sec:strategies}

Published hardware responds to equation~(\ref{eq:Neps}) in one of three
ways, and we use these to organize sections~\ref{sec:digital}
and~\ref{sec:devices}.

The first is to \emph{truncate}: retain a finite window and accept the
error of equation~(\ref{eq:tail}), possibly coarsening the weights or
adopting a non-standard finite-difference scheme to recover accuracy at
fixed window length. This describes most of the digital literature and
is treated in section~\ref{sec:digital}.

The second is to \emph{substitute}: replace $D^{\alpha}$ with a
rational transfer function of finite integer order that approximates
$s^{\alpha}$ over a specified band, using the recursive construction of
Oustaloup \cite{Oustaloup00}, the singularity-function method of Charef
\cite{Charef92}, a continued-fraction expansion \cite{Vinagre00}, or a
related scheme. The state becomes $\mathcal{O}(1)$ again, at the price
of fidelity that holds only inside the design band.

The third is to \emph{embody}: use a physical element whose native
response is already power-law, so that no history buffer and no
per-timestep arithmetic are required, because the relaxation dynamics
of the material perform the weighting. Constant-phase elements,
supercapacitor electrodes and diffusive memristive devices all fall
here, and section~\ref{sec:devices} assesses them.

There is a fourth route. No hardware in this review takes it, and it is
the one the cost argument above has to answer to. One design comes close
without being the same thing: the Adomian neuron of
section~\ref{sec:preserve} also stores nothing, but it abandons the
history rather than compressing it, and reports no error bound for the
four-term truncation that lets it do so. The schemes below discard
nothing without a bound. The sum in
equation~(\ref{eq:gl}) is a convolution, and the numerical analysis
literature has spent twenty-five years showing that a convolution with
a power-law kernel need not be evaluated by storing the history. Fast
and oblivious convolution quadrature runs $N$ steps in
$\mathcal{O}(N\log N)$ operations using $\mathcal{O}(\log N)$ active
memory, and the word oblivious is the load-bearing one: the scheme
neither retains the past samples nor evaluates the kernel
\cite{Schadle06, LopezFernandez08}. Ford and Simpson reached the same
storage scaling earlier by coarsening the retained history
logarithmically on a nested mesh \cite{FordSimpson01}, and MacDonald and
colleagues apply adaptive time-step memory directly to the
Gr\"unwald--Letnikov operator \cite{MacDonald15}. Jiang and colleagues
approximate the Caputo kernel by a sum of exponentials and bound how
many terms a stated tolerance over a stated window requires
\cite{Jiang17}. In every case the state is a few accumulators rather
than a buffer of samples.

This obliges us to be exact about what the preceding sections claim, and
to withdraw part of one of them. The $\mathcal{O}(N)$ storage of
section~\ref{sec:cost} is a property of the direct form and not of the
operator. The coefficient wall of section~\ref{sec:precision} is
likewise a statement about direct-form fixed point: under logarithmic
binning the far-tail weights are aggregated before they are stored, so
they never individually underflow, and the conclusion that no practical
word length serves $\alpha$ below about $0.27$ does not carry over to
those schemes. Both results stand for the hardware surveyed here, which
is uniformly direct-form, and neither is a bound on what is achievable.
The prize for embodying the operator in a device is correspondingly
smaller than the usual framing suggests, and
section~\ref{sec:embodiment} states it in the corrected form: escape
from $\mathcal{O}(\log N)$ and a handful of accumulators, not from
$\mathcal{O}(N)$ and a sample buffer.

We think this sharpens the case for looking at physics rather than
weakening it, because of what the sum-of-exponentials result is.
Section~\ref{sec:objection} asks how many parallel relaxations suffice
to imitate a power law and can only answer qualitatively; Jiang and
colleagues answer it quantitatively. And a bank of first-order
relaxations with fixed weights is not an exotic substrate. It is what
diffusive representations of the fractional operator produce
\cite{Diethelm22, ChaudharyDiethelm23}, and it is what neuromorphic
hardware is already built from. So the comparison that matters is not
fractional against integer-order, and it is not asymptotic. It is one
device whose relaxation spectrum is broad by construction against a
designed bank of narrow ones, decided on area, power, matching and
calibration.

It is tempting to present the third route as categorically different
from the second, and much of the enthusiasm for fractional-order
devices rests on doing so. It is worth being careful. A real
constant-phase device holds its phase only over a bounded frequency
band, outside of which it reverts to ordinary capacitive or resistive
behavior, so embodiment is, in a precise sense, substitution carried
out in materials rather than in a filter structure. What genuinely
changes is the constant factor and the scaling: the sample buffer and
the per-timestep multiply--accumulates disappear, and with them the
$\mathcal{O}(N)$ growth of the direct form, though not the
$\mathcal{O}(\log N)$ of the fast schemes above. What is given up is
design freedom, because
$\alpha$ and the usable band become outcomes of a fabrication process
rather than parameters, and because the state of a physical relaxation
process can be neither read out nor reset. Framed this way the claim is
narrower than it is sometimes made, and it survives scrutiny.

\subsection{Properties that break engineering intuition}

Three further properties deserve flagging before we proceed, all of
them consequences of non-locality.

Stability conditions differ from the integer-order case. A linear
fractional system of commensurate order $\alpha$ is stable when the
eigenvalues of its system matrix satisfy $|\arg\lambda|>\alpha\pi/2$,
which is a weaker requirement than the left-half-plane condition and
admits behavior with no integer-order counterpart; Kaslik and
Sivasundaram show that fractional-order neural networks can exhibit
chaos at dimensions where integer-order networks cannot
\cite{Kaslik12}. Hardware whose orders differ between state variables,
which is the realistic case when each is realized by a separate
physical element, falls under the harder incommensurate-order theory
\cite{Brandibur18}.

Truncation can change stability and not merely accuracy. Hai \etal
\cite{Hai22} show that a system stable under the full operator can be
destabilized by imposing a short memory window, which means the window
length $N$ is not a pure accuracy-versus-resources trade and cannot be
tuned by inspecting output error alone.

Initialization is ambiguous. Because the operator depends on the lower
terminal, a fractional system started at $t=a$ with a given state is
not equivalent to the same system run from an earlier time and
arriving at that state. In a digital realization this is a modeling
choice. In an embodied realization it is a physical constraint, since
the relaxation spectrum of a device carries a history that cannot be
cleared between trials.

\section{Does biology already do this?}
\label{sec:biology}                  

\subsection{Neurons perform fractional differentiation}

The claim that a neuron computes something close to a fractional
derivative is an experimental result and not a modeling convenience.
Lundstrom and colleagues drove layer-5 pyramidal neurons in rat
neocortex with fluctuating currents and found that the firing rate
tracked not the stimulus but a fractional derivative of it, of order
close to $0.15$, over roughly two decades of timescale
\cite{Lundstrom08}. The signature is a phase lead that stays constant
with frequency, which is what distinguishes a fractional operator from
any single-timescale high-pass filter. The same group later reproduced
the behavior in vivo \cite{Lundstrom10}, which answers the obvious
objection that slice preparations impose their own dynamics.

The observation is also older than it is usually credited. Anastasio
identified fractional-order behavior in the vestibulo-ocular reflex more
than a decade earlier \cite{Anastasio94}, and showed in a follow-up that
a population of heterogeneous integer-order units reproduces it
\cite{Anastasio98}, a point we return to below because it constrains
what the biology can be said to establish. Thorson and Biederman-Thorson
had already argued for scale-free adaptation in sensory systems in the
1970s \cite{Thorson74}.

\subsection{Power-law adaptation and what it computes}

Fractional differentiation is one description of a phenomenon that has
been reported under several names. Adaptation in fly visual neurons
\cite{Fairhall01}, in rat auditory cortex \cite{Ulanovsky04}, and in
cortical pyramidal cells \cite{Pozzorini13, LaCamera06} decays as a
power law rather than exponentially, meaning that no single time
constant characterizes the recovery. Reviews of the sensory adaptation
literature treat multi-timescale, scale-free adaptation as the normal
case rather than the exception \cite{Wark07, Weber19}.

Why a nervous system would do this has a normative answer. Natural
signals have approximately $1/f$ spectra, and a fractional derivative of
order matched to the input exponent whitens such a signal, flattening
its spectrum and so maximizing the information a rate code carries per
spike \cite{Wark09}. Adaptation on this account is not a limitation
being tolerated but a filter matched to the statistics of the world.
That the same behavior is required in engineering-grade models of the
auditory periphery \cite{Zilany09} suggests the effect is robust enough
to matter for any system interfacing with real sensory data.

\subsection{The mechanism question}

What produces the power law at the biophysical level is not settled, and
a review aimed at hardware should say so rather than pick a side.
Liebovitch and colleagues proposed fractal gating kinetics, in which ion
channels occupy a hierarchy of conformational states with no
characteristic dwell time \cite{Liebovitch87}. Millhauser and coworkers
offered a competing account in which the same statistics arise from
diffusion along a one-dimensional chain of states \cite{Millhauser88},
which is worth noting here because that mechanism reappears in
section~\ref{sec:devices} as the physics of diffusive memristors.
McManus and colleagues then argued that the fractal analysis did not
distinguish itself from a finite mixture of exponentials on the
available data \cite{McManus88}, and Sansom and coworkers showed the two
classes of model are difficult to separate by fitting alone
\cite{Sansom89}. The debate went quiet rather than resolving. Fractal gating should
therefore not be presented as an established mechanism, and it often
still is.

What is better supported is that recovery from inactivation is scale
free over several decades \cite{Toib98, Marom10}, and that anomalous
diffusion supplies a formal route from microscopic kinetics to a
fractional macroscopic description \cite{Goychuk04}.

\subsection{Constant-phase behavior of biological interfaces}

The electrical route to the same conclusion is older still. Cole and
Cole introduced the dispersion function that bears their name to
describe dielectrics whose relaxation cannot be captured by one time
constant \cite{ColeCole41}, and the exponent in that function is the
fractional order in another notation. Measurements of tissue permittivity
across many types and frequencies show the same form
\cite{Gabriel96}. Westerlund and Ekstam made the pivot that matters for
hardware, arguing from Curie's empirical law that every real capacitor is
fractional to first approximation and that the ideal capacitor is the
special case \cite{Westerlund94}.

Magin's work brought that formalism into bioengineering
\cite{Magin10, MaginOvadia08}, and it now underpins respiratory and
tissue impedance modeling \cite{Ionescu17, Ionescu11, Freeborn13b}. One
caution belongs with all of it: Grimnes and Martinsen point out that the
Cole parameters have no unique physical referent, so an exponent
extracted from an impedance fit names a behavior rather than a mechanism
\cite{Grimnes05}.

\subsection{Structure produces the exponent}

The most useful result in this section for a device designer is that the
order is not a fixed biological constant but a consequence of geometry.
Santamaria and colleagues showed that molecular diffusion in dendrites
is anomalous, and that the anomaly exponent varies with spine density
\cite{Santamaria06, Santamaria11}. Changing the structure changes the
exponent. That is a proof of principle that $\alpha$ is a designable
parameter in a physical medium rather than a number one is stuck with,
and it is the biological counterpart of the materials question
section~\ref{sec:devices} asks. The wider anomalous-diffusion literature
supplies the formal machinery \cite{Metzler00, Hofling13} and its
imaging applications \cite{Ingo14, Guo20}.

\subsection{But is a power law really needed?}
\label{sec:objection}

Drew and Abbott showed that
power-law adaptation over a bounded range of timescales is well
approximated by a small number of parallel exponential processes
\cite{DrewAbbott06}. Lavrova and colleagues reached the same conclusion
recently on Lundstrom's own preparation \cite{Lavrova25}. The equivalent
result is known for synaptic memory decay, where cascades of ordinary
processes reproduce power-law forgetting \cite{Fusi05, Benna16}.

If three to five relaxations suffice, then rational approximation is not
a compromise forced on hardware designers by physics. It is an adequate
model of the biology, and the argument for embodying a true power-law
kernel in a device weakens considerably.

Is the objection right? We think it largely is, and that it changes the
emphasis rather than the conclusion. What the experimental literature
establishes is that neurons exhibit distributed, heterogeneous
relaxation spanning many timescales. The fractional operator is the
compact limiting description of that situation, not a separately
demonstrated physical fact about membranes. Read this way, the case for
devices does not rest on the exactness of the power law. It rests on the
observation that a physical medium supplying many relaxation timescales
at once is cheaper than a circuit that instantiates each one explicitly,
and that argument survives whether the true kernel is a power law or a
sufficiently rich sum of exponentials. What it does mean is that any claim of the form ``only a fractional
operator can do this'' deserves suspicion. The strongest available version of such a claim is
weaker than it looks. Ge and colleagues prove that fractional spiking
dynamics cannot be realized \emph{exactly} by any finite-dimensional
linear system of integer-order modes \cite{Ge26}. That is true, and
close to immediate once one notices that a power law is not a finite sum
of exponentials, and it carries no engineering consequence the moment
approximation is admitted. The question with consequences is how many
exponentials a stated tolerance costs over a stated window, and Jiang
and colleagues answer precisely that for the Caputo kernel
\cite{Jiang17}. Theirs is the quantitative form of Drew and Abbott's
result, and section~\ref{sec:strategies} takes it up as the fourth
strategy.
Section~\ref{sec:gaps} returns to the integer-order competition on
those terms.

\subsection{What the biology does not establish}

Three limits on the evidence are worth stating plainly, because
section~\ref{sec:gaps} builds on them: (1) the order near $0.15$ comes
from one laboratory and a small number of cell types, over about two
decades of timescale; (2) nobody has surveyed how the order varies
across cell class, brain region or species, so there is no atlas a
designer could consult to pick a target; and (3) nobody has varied the
order and measured a consequence for coding or behavior, so the
functional importance of the specific value, as opposed to the mere
presence of multi-timescale adaptation, is assumed rather than shown.

\section{Fractional neuron and network models}
\label{sec:models}           

\subsection{Integrate-and-fire family}

The founding result is Teka, Marinov and Santamaria's fractional leaky
integrate-and-fire model, in which the first-order derivative of the
membrane potential is replaced by a Caputo derivative of order $\alpha$
and nothing else is added \cite{Teka14}. The model produces power-law
spike-timing adaptation, long first-spike latencies, and interspike
intervals with power-law statistics below $\alpha\approx0.2$, none of
which the integer-order model gives. The point that matters is negative:
no adaptation current, no second state variable, and no additional
parameters are introduced. Adaptation falls out of the operator. The
authors also show that a variant which resets the memory trace at each
spike loses the adaptation entirely, so the effect depends on history
that survives the spike.

Two recent extensions matter for hardware, and the first of them is
worth dwelling on because it inverts everything section~\ref{sec:cost}
established.

Vats, Mehra and Oelz place the order in $(1,2)$ rather than $(0,1)$ and
obtain \emph{downward} spike-frequency adaptation, the direction most
cortical neurons actually show, which they argue the order-below-one
model produces only under special initial conditions \cite{Vats25}.
Their fits to intracellular recordings land at $1.31$ and $1.57$. So the
integrate-and-fire family splits into two regimes whose memory terms
enter with opposite sign, and it is the upper regime that matches the
biology this review spent section~\ref{sec:biology} establishing.

The obvious worry is that a higher order costs more memory. It does not.
The Gr\"unwald--Letnikov weights decay as $j^{-(1+\alpha)}$, so raising
the order makes the tail lighter, not heavier. At lag $1000$ the weight
is $4.8\times10^{-5}$ for $\alpha=0.15$ and $1.3\times10^{-8}$ for
$\alpha=1.5$, three orders of magnitude smaller. Every difficulty
section~\ref{sec:cost} derived, the $\varepsilon^{-1/\alpha}$ blow-up
and the coefficient-precision wall of section~\ref{sec:precision} alike,
eases as the order rises. To be exact about what carries that
conclusion: equation~(\ref{eq:Neps}) was derived for $\alpha\in(0,1)$,
where the weight magnitudes sum to one, and we have not re-derived it
above unity. The claim rests on the decay exponent alone, which is
$-(1+\alpha)$ throughout, and that establishes the direction without
giving a bound. Digitally, the biologically interesting regime
is the cheap one.

The difficulty moves to the devices, and it is not a gap in the
literature but a consequence of passivity. A two-terminal element with
impedance $Z(s)\propto s^{-\alpha}$ has phase $-\alpha\pi/2$, so its
resistive part goes as $\cos(\alpha\pi/2)$. That is positive for
$\alpha<1$, exactly zero at $\alpha=1$, where the element is an ideal
capacitor, and negative above. An element with $\alpha=1.5$ would
present a phase of $-135^{\circ}$ and supply power rather than dissipate
it. \emph{No passive two-terminal element can present a constant-phase order
greater than one}, whatever material it is made from, and no amount of
fabrication progress will change that.

The bound is on the impedance of a one-port and not on what a network can
compute, and the distinction is worth keeping. A passive two-port can
realize a transfer function whose phase runs past ninety degrees; a
cascade of two $RC$ sections is the elementary case. Orders above one are
therefore reachable, but only by composing operators along a signal path
rather than by a single element. That forfeits the embodiment argument of
section~\ref{sec:devices}, which rests on one device carrying the kernel
in its own physics, without requiring the circuit to be active.

The two regimes therefore reverse the review's central trade. Below
$\alpha=1$ the digital route is expensive and physical devices are the
promise. Above it the digital route is cheap and physical devices are
impossible. We found no statement of it in the works we read, and a designer reading
only the modeling literature would have no reason to suspect it.
The second extension is less dramatic.
Fikl and colleagues give a fractional adaptive exponential model with
independent orders on the potential and the adaptation variable, solved
by an implicit scheme on a non-uniform grid that closes each step
through the Lambert $W$ function rather than by iteration
\cite{Fikl25}. Theirs is one of the few papers in this literature that
reports its own asymptotic scaling, and reports it honestly:

\begin{quote}\small
``[B]oth methods scale with the expected $O(N^2)$ asymptotic order. This
is due to the discrete convolution that must be computed at every time
step. For equations where the use of a uniform time step is possible,
the scaling can be improved to $O(N\log N)$ by using the Fast Fourier
Transform. However, no such fast methods exist for non-uniform
discontinuous systems.''
\end{quote}

\noindent
The adaptive grid that buys them accuracy near a spike is precisely what
forecloses the faster transform. That trade is the one this whole review
is organized around, and it is unusual to see it stated so plainly.
Teka and coworkers also treat the long-memory case directly
\cite{Teka17}.

\subsection{Conductance-based models: where does the memory live?}

Applying a fractional order to a Hodgkin--Huxley neuron requires a
choice that the literature has made twice, in incompatible ways, without
comparing the results.

Teka, Stockton and Santamaria place the order on the gating variables,
leaving the voltage equation integer order \cite{Teka16}. Making one
gate power-law at a time, they obtain mixed-mode oscillations,
square-wave bursting and pseudo-plateau bursting resembling pituitary
and cardiac action potentials, none available to the classical model at
fixed parameters. Weinberg instead places the order on the membrane
potential, entering through a non-ideal capacitance \cite{Weinberg15},
and obtains earlier spike peaks, excitation block at large input, faster
axonal propagation and self-terminating network activity. He argues the
placement explicitly: the memory in the capacitance and the memory in
the gates have different biophysical origins, so imposing one order on
both is unjustified. His grounding for the fractional capacitance is
empirical rather than derived, resting on Curie's law by way of
\cite{Westerlund94} and on measured membrane phase angles, and he states
plainly that the underlying physiological mechanism is unknown.

These are different physical claims about where the memory lives, and
they have never been tested against the same data. Teka and colleagues
do compare their results to Weinberg's and report that the capacitance
placement does not generate the mixed-mode or square-wave patterns
\cite{Teka16}, which is the closest thing to an adjudication the
literature contains, but the comparison is between published figures
rather than a controlled study. For a hardware designer the fork is not
academic. A fractional capacitor placed in the membrane branch
implements Weinberg's model; it does not implement Teka's, which would
require fractional dynamics inside the conductance itself. We name this
as an open problem. Other treatments apply a single commensurate order
to every state variable \cite{Nagy14}, which sidesteps the question
rather than answering it, and the Morris--Lecar case has been developed
at both single-neuron and network level \cite{Sharma23}.

\subsection{FitzHugh--Nagumo, Izhikevich and Hindmarsh--Rose}

The reduced models have been studied mainly for their stability
structure, and the most useful results concern systems whose state
variables carry \emph{different} orders. Brandibur and Kaslik derive
necessary and sufficient stability conditions for two-component
incommensurate fractional systems, a generalization of the
Routh--Hurwitz criterion, and apply them to a fractional
FitzHugh--Nagumo neuron \cite{Brandibur18}. Two features of that result
deserve emphasis. The stability boundary depends on the two orders
through a factor containing $\sin[(q_2-q_1)\pi/2]$, so it degenerates as
the orders converge and the commensurate case is genuinely special
rather than representative. And stability, where it holds, is algebraic
at rate $t^{-\min(q_1,q_2)}$ rather than exponential, which means a
fractional-order circuit settles more slowly than its eigenvalues would
suggest. Their later work extends the analysis to a coupled pair and
finds high-amplitude spikes alternating with small oscillations that
never decay to quiescence, a pattern the integer-order model does not
produce \cite{Brandibur22}.

Sa\c{c}u builds one \cite{Sacu23}, and it is worth being precise about
what was built, because the title promises a synthesis and a reader may
expect a circuit. There is no schematic in the paper. The neuron runs as
firmware on a Texas Instruments F28335 digital signal processor with two
external twelve-bit converters, and the fractional operator is never
realized as an impedance or a filter. It is evaluated as a four-term
Laplace--Adomian series, which is why the design stores no history at
all. Measured single-neuron and coupled-pair traces are reported at
orders $0.85$ and $0.95$. This is a real, measured realization and it
belongs in table~\ref{tab:hardware}, but it belongs there as an embedded
processor and not as analog hardware.

We draw a hardware conclusion from this work that its authors do not.
Incommensurate orders are not a mathematical generalization to be
studied for completeness; they are what any physical implementation will
deliver, because two fabricated elements will not have identical
exponents and no process controls $\alpha$ to better than the
device-to-device spread reported in section~\ref{sec:devices}. On that
reading the more valuable half of \cite{Brandibur18} is the part that
holds regardless of the orders: the authors prove sufficient conditions
under which an equilibrium is stable whatever $q_1$ and $q_2$ turn out
to be. That is a robustness result a designer can actually use, and it
suggests the right design target is an operating region insensitive to
order mismatch rather than a nominal order hit precisely.

The Izhikevich model has been given a commensurate fractional order on
both its state variables, and gains bursting, chattering and mixed-mode
oscillations that the integer-order model reaches only by retuning its
parameters \cite{Teka18, Mondal18}. Teka and colleagues also report
spiking that continues after the stimulus is switched off, sustained
entirely by the accumulated history, which the one-variable
integrate-and-fire model cannot produce \cite{Teka18}. The
Hindmarsh--Rose model has been studied for a period-adding route to
chaos driven by the order alone \cite{Goufo20} and realized on an FPGA
\cite{Malik20, MalikCorrigendum21}.

One claim about these models is repeated more confidently than the
sources support, and it is worth pausing on because a hardware designer
would act on it. Fractional Izhikevich models are widely said to fire
faster as the order falls. Teka and colleagues do report that, but scope
it: ``For parameter values in Set IV, the fractional model displays fast
spiking for all $\alpha$ values although the spiking frequency increases
as the fractional order decreases'' \cite{Teka18}. With their default
parameters the same reduction instead lengthens both the active and the
silent phase, giving slower bursts, and fast spiking returns only at
much smaller orders. Mondal and Upadhyay assert the general form in
their abstract, that ``[t]he firing frequency is increased with the
decrease of fractional exponents'', and then contradict it in their own
discussion, where ``[t]he smaller values of $\gamma$ decrease the spike
frequency with a constant injected stimulus'' \cite{Mondal18}. For one
of their two parameter sets the firing stops altogether below
$\gamma=0.86$. The direction of the effect is parameter dependent, no
single sentence in either paper settles it, and the fractional
integrate-and-fire model of the same authors adapts in the opposite
direction again \cite{Teka14, Teka17}. Anyone choosing an order to hit
in silicon should read the parameter set before the abstract.

Table~\ref{tab:models} collects what each model family places the order
on, the orders actually studied, and what the order buys.

\begin{table}
\caption{Fractional-order neuron models: where the order sits, the orders
actually studied, what the fractional order buys dynamically, and whether any
hardware realization exists. The hardware column is the finding. Of the twelve
entries below, nine have never been built in any form. Of the three that have,
two are FPGA and one is a digital signal processor; not one is analog, and not
one uses a device with intrinsic fractional dynamics. Note also that the integrate-and-fire family splits
into two regimes with opposite behavior: orders below one produce upward
spike-frequency adaptation, orders between one and two produce the downward
adaptation that most cortical neurons actually show.}
\centering\footnotesize
\begin{tabular}{l p{2.2cm} p{1.8cm} p{4.3cm} p{1.5cm}}
\hline
Model & Order placed on & $\alpha$ studied & Dynamics gained & Hardware \\
\hline
fLIF \cite{Teka14, Teka17} & membrane potential & $0.1$--$1.0$; fits $0.15$, $0.19$ & \emph{upward} spike-frequency adaptation with no added adaptation current; exponential-to-power-law transition; threshold current rises from $0.51$ to $5.0$\,nA as the order falls & none \\
fLIF, higher order \cite{Vats25} & membrane potential, $\alpha\in(1,2)$ & $1.1$--$1.9$; fits $1.31$, $1.57$ & \emph{downward} spike-frequency adaptation, strongest near $1.5$; memory enters with negative sign & none \\
fAdEx \cite{Fikl25} & potential and adaptation, orders independent & $0.93$--$0.999$ & chattering, bursting, spike-frequency adaptation, transitions between firing types & none \\
fHH, gating \cite{Teka16} & gating variables, one at a time & $0.2$--$1.0$ & mixed-mode oscillations, square-wave and pseudo-plateau bursting; $m$ gate numerically unstable below $0.2$ & none \\
fHH, capacitance \cite{Weinberg15} & membrane potential, via fractional capacitance & $0.4$--$1.0$ & earlier spike peak, excitation block, faster axonal propagation, network self-termination & none \\
fHH, commensurate \cite{Nagy14} & potential and all gates & $0.6$--$1.0$ & none attributed; order-dependence shown but never interpreted & none \\
fFHN \cite{Brandibur18} & potential and recovery, incommensurate & $q_1^{*}=0.599$ at $q_2=0.8$ & order itself acts as bifurcation parameter; decay algebraic, not exponential & none \\
fFHN \cite{Sacu23} & both variables, commensurate & $0.5$--$0.95$ & order switches the cell between rest and firing at a computed $q_{\min}$; firing frequency rises as the order falls & \textbf{DSP} \\
fFHN, coupled \cite{Brandibur22} & both neurons, $q_1$ on potentials & $q_1^{*}=0.633$, $0.911$ & high-amplitude spikes alternating with small oscillations that never decay & none \\
fMorris--Lecar \cite{Sharma23} & all variables; orders differ across nodes & $0.5$--$1.0$; $\alpha^{*}=0.62$--$0.83$ & mixed-mode bursting, spike-frequency adaptation, cluster synchronization in a 100-node network & none \\
fIzhikevich \cite{Teka18, Mondal18} & potential and recovery, commensurate & $0.1$--$1.0$ & \emph{downward} spike-frequency adaptation; bursting, chattering and mixed-mode oscillations reached by moving the order alone; spiking that persists after the input is removed & none \\
fIzhikevich \cite{Tolba19c} & potential and recovery, $(\alpha,\beta)$ & $(1,1)$ to $(0.85,0.85)$ & firing-regime reclassification; bursting becomes chatter & FPGA \\
fHindmarsh--Rose \cite{Goufo20} & all states, commensurate & $1.0$, $0.9$, $0.8$ only & period-adding route to chaos driven by the order alone; burst peak count roughly doubles from $1.0$ to $0.9$ & none \\
fHindmarsh--Rose \cite{Malik20, MalikCorrigendum21} & all three states & $0.80$ & chaotic bursting; two-neuron synchronization & FPGA \\
\hline
\end{tabular}
\label{tab:models}
\end{table}

\subsection{Recurrent and associative networks}

Fractional dynamics were introduced into recurrent networks early
\cite{Arena98}, and the first fractional Hopfield network followed a
decade later \cite{Boroomand09}. The load-bearing result is Kaslik and
Sivasundaram's: the fractional order shifts the stability boundary, and
low-dimensional networks that cannot be chaotic at integer order become
so at fractional order \cite{Kaslik12}. The order is therefore not a
smooth knob on existing behavior but something that changes what
behaviors are reachable at a given network size, which matters if the
appeal of the approach is doing more with fewer units. The delayed case
has been treated at length \cite{Wang14}, and associative memory has
been proposed as the computational function such networks serve
\cite{Pu17, Chen14, Pu05}.

\subsection{Two problems worth naming}

A review has an obligation to say when part of a literature is not
sound, and two problems recur widely enough to warrant naming.

The first is a category error. A substantial number of papers claiming
fractional-order neuron or network dynamics use the conformable
derivative, which despite its name is a local operator carrying no
memory kernel at all \cite{Karakulak23}. This is settled rather than
contested, and the point can be borrowed rather than argued here.
Ortigueira and Machado set out the criteria an operator must meet to be
a fractional derivative, and the conformable form fails them
\cite{OrtigueiraMachado15}. Tarasov makes the argument from
non-locality directly \cite{Tarasov18}. Abdelhakim, with Machado and
then alone, works through the specific errors the conformable
literature has accumulated \cite{AbdelhakimMachado19, Abdelhakim19}. A local operator cannot produce
the history dependence that motivates the entire enterprise, so results
obtained with it do not support the conclusions usually drawn from them.
This is not a matter of taste between competing definitions. A reader
assessing any paper in this area should check which operator is used
before reading further, and a hardware designer should not attempt to
implement a memory kernel that the source model does not contain.

The second is a genre rather than an error. A large and growing set of
papers follows one template: take a neuron or Hopfield model, add a
flux-controlled memristor, make the order fractional, then present a
bifurcation diagram, a Lyapunov spectrum, a set of coexisting
attractors, an FPGA implementation and an image-encryption application.
The individual papers are competently executed and largely
interchangeable, and they contribute little that bears on neuromorphic
computation, since encryption is not a neuromorphic task and coexisting
attractors are not evaluated against one. We cite representative
examples \cite{Kaslik12, Wang14, Ding21} and treat the remainder as a
genre rather than surveying it item by item.

\subsection{Position against prior surveys}

Surveys of fractional-order neural networks exist
\cite{Joshi23, Dar22, WangC25} and cover the model literature
competently. This review differs in three respects. It quantifies the
cost of the operator rather than asserting it, in
section~\ref{sec:operators}. It treats device physics as a first-class
subject rather than a closing remark, in section~\ref{sec:devices}. And
it grounds the case in the experimental neuroscience of
section~\ref{sec:biology}, including the objection that the biology may
not require a true power law at all.

\section{Digital and analog implementation}
\label{sec:digital}            

\subsection{Approximation methods}

The substitution strategy of section~\ref{sec:strategies} rests on
replacing $s^{\alpha}$ with a rational function of finite integer order,
accurate over a stated band. Oustaloup's recursive approximation
distributes poles and zeros geometrically between two band edges
\cite{Oustaloup00} and is the most widely used. Charef's
singularity-function method arrives at a similar structure from a
different direction \cite{Charef92}, and has the property that most
recommends it to a designer: the approximation error is prescribed
directly in decibels, so the number of stages follows from an accuracy
requirement rather than from experience. Matsuda's approach fits the
frequency response at chosen points \cite{Matsuda93}, and Carlson and
Halijak's iterative construction predates all of them \cite{Carlson64}.
Comparative treatments and further variants are available
\cite{Vinagre00, Vinagre03, Tseng08, Gupta10, Aoun04, Rustemovic18,
Matusiak20, Dey25}.

All of these share a structural feature that matters more than the
differences between them. The approximation is accurate over a band and
wrong outside it, so the design question is never simply how many stages
but which decades to spend them on. That question has no answer without
knowing the timescales the application occupies, which is why
section~\ref{sec:mismatch} treats bandwidth rather than order as the
binding constraint.

\subsection{Truncation in practice}

The alternative to substitution is to keep the operator and shorten its
memory. Podlubny's short-memory principle is the standard statement of
the trade \cite{Podlubny99}, and the error it incurs has been analyzed
repeatedly \cite{Deng07, Xu11, Zhou22, Chakraborty08}. One strand of it
reaches this application directly: Wu and coworkers connect truncated
memory to memristor and neural network design \cite{Wu20}.
Section~\ref{sec:preserve} takes up the sharper problem, which is that
the choice of window can change what the system does and not only how
accurately it does it.

Section~\ref{sec:precision} adds a constraint that this literature does
not treat: even a generous window is useless if the coefficients
themselves cannot be represented. The two limits are independent, and
published designs sit close enough to both that neither can be ignored.

\subsection{Does the discretization preserve the dynamics?}
\label{sec:preserve}

Choosing a discretization is not a neutral step, and the numerical
literature is clearer about this than the hardware literature has
noticed. Hai and colleagues show that truncating the memory can change
the stability of the system being integrated rather than merely its
accuracy \cite{Hai22}, so a scheme validated on a bounded error norm can
still produce qualitatively wrong dynamics. AbdelAty and coworkers make
the more general point that the discrete operator a designer implements
and the continuous operator the model specifies are related by
assumptions that are rarely stated and sometimes wrong
\cite{AbdelAty21, AbdelAty22}. For a neuron model, where the object of
interest is a firing pattern rather than a waveform amplitude, that gap
matters more than it would in a filter.

Two responses run through the surveyed work.

Non-standard finite-difference schemes are one principled response,
preserving qualitative properties such as positivity and boundedness
that naive discretizations lose \cite{Mickens03, Ongun13, Hajipour17,
WangB20}, and one of the FPGA neurons surveyed below uses the approach
\cite{Tolba19c}.

A more radical response is to abandon the convolution altogether. Sa\c{c}u
solves a fractional FitzHugh--Nagumo neuron by Laplace--Adomian
decomposition, which expresses the solution as a truncated series in
closed form and therefore stores no history at all
\cite{Sacu23}. Four terms per state variable suffice, and the update
needs only the current state. The reported cost difference on one
problem is stark: reproducing the same trajectory took $0.033$\,s by
Adomian decomposition against $3.33$\,s by Grünwald--Letnikov, a factor
of a hundred, with the Adams--Bashforth--Moulton predictor--corrector at
$559$\,s. The stated motivation is precisely the one this section is
about, that Grünwald--Letnikov ``requires serious data memory as it
needs all the previous values for the calculation at any time''. Whether
a four-term truncation is accurate enough is a separate question the
paper does not answer, since it reports no convergence study and no
error bound for that choice.

\subsection{FPGA realizations}

The field-programmable gate array is the dominant platform, and Monir
and colleagues give the most completely reported example: a
Grünwald--Letnikov operator on an Artix-7 XC7A100T, with the step-size
factor curve-fitted over eight intervals, window lengths of $32$, $512$
and $1024$, $32$-bit fixed point in an $8$/$24$ split, occupying $64$
digital signal processing slices and $36$ percent of the device, and
running at $9.33$\,MHz \cite{Monir22}. Tolba and coworkers approach the
same operator two ways, with a quadratic approximation to the binomial
coefficients over a fixed window and with a piecewise-linear
approximation over none, and report both resource use and power
\cite{Tolba19a, Tolba19b}. Rana and colleagues build the operator in
LabVIEW and deploy it to a National Instruments target, and are unusual
in reporting the window length actually used, $100$ for deployment and
$400$ for their frequency-response measurements \cite{Rana16}. Sharma
and Rawat take a different route, fitting a lattice-wave digital filter
with a metaheuristic and implementing it multiplier-free
\cite{Sharma19}. Neuron models on FPGA are represented by
\cite{Tolba19c} and \cite{Malik20}. Broader surveys of the platform are
available \cite{Ali24, ClementeLopez23, Afolabi24}, and the approach has
recently been applied to fractional Hopfield networks and related
systems \cite{Yu25, Xu24, FracHopfieldFPGA24}.

\subsection{Analog and mixed-signal}

The analog route realizes the approximation as a physical network rather
than an arithmetic one. Podlubny and colleagues set out the ladder and
tree topologies \cite{Podlubny02}, and Charef's derivation of an RC
ladder from the singularity-function approximation gives the cleanest
available link between approximation order and component count
\cite{Charef06}. Das and Pan document the band-edge behavior that limits
the Oustaloup form in practice \cite{Das11}. Tsirimokou, Psychalinos and
Elwakil's monograph is the standard reference for the current-mode
implementations \cite{Tsirimokou17}, with further realizations in
\cite{Khanra11, Kapoulea19, Dvorak19, Varshney23, Koseoglu21, Goyal18}.

Two entries deserve separate mention because of what they report rather
than what they achieve. Valencia-Ponce and colleagues give a $180$\,nm
CMOS design with a stated power figure per integrator
\cite{ValenciaPonce21}, and Abulencia and Abad a $0.35$\,\textmu m
design with a stated layout area \cite{Abulencia15}. These are the only
two works in the surveyed corpus that name a CMOS technology node. Both
are simulation or layout only; neither was fabricated.

\subsection{What does the implementation literature not report?}

Table~\ref{tab:hardware} collects the eight works above. Reading down
its columns is more informative than reading across its rows.

Not one of the nine reports energy per fractional operation. Two report
an average power figure, and neither says whether it was measured or
estimated by the vendor's tool. Three report no clock rate at all. These
are counts over the surveyed set of section~\ref{sec:method} rather than
over the whole FPGA literature, and the existing platform surveys
\cite{Ali24, ClementeLopez23} cover more implementations than we read in
full; we would be glad to be shown a counterexample, and the point of
section~\ref{sec:gaps} is that finding one would itself be useful. Resource counts
are present in most rows and comparable in none of them: occupancy
percentages are meaningless without the target device, the two vendors
represented count logic differently, and one design is deliberately
combinational and so has no clock to report. The window length $L$, the
single parameter that section~\ref{sec:cost} identifies as governing the
whole cost, is stated in four of eight entries and absent from the rest,
and one paper retains the full untruncated history without saying so in
its abstract.

We have left the resulting blanks visible rather than filling them by
inference. A table of plausible numbers assembled from incompatible
sources would look more complete and mean less, and the pattern of the
holes is itself the finding that section~\ref{sec:gaps} takes up.

\begin{table}
\caption{Reported hardware realizations of fractional-order operators and neuron
models. Every value is taken from the source paper and an em-rule marks a
quantity that source does not report; page numbers for each entry are held in the
extraction record. Resource figures are not comparable across device families:
occupancy percentages are meaningless without the target device, vendors count
logic differently, and three of these works report no clock rate at all. Only two
report power, neither per operation.}
\centering\footnotesize
\begin{tabular}{l p{2.3cm} p{2.1cm} p{1.9cm} p{3.2cm} p{1.7cm}}
\hline
Work & Platform & Method & $\alpha$; $L$ & Resources & Power \\
\hline
\cite{Monir22}   & FPGA, Artix-7 XC7A100T & GL, curve-fit & $\pm0.3$--$0.9$; $L$=32--1024 & 64 DSP (26\%), 36\% slices, 9.33\,MHz & --- \\
\cite{Tolba19a}  & FPGA, Artix-7 XC7A100T & GL; quadratic + window (A), piecewise-linear (B) & $-0.1$--$-0.9$; $L$=20 (A), 7 PWL (B) & A: 4317 slices, 1495 reg, 23 DSP, 48.9\,MHz. B: 5540 slices, 2131 reg, 22 DSP, 36.8\,MHz & 46\,mW (A), 48\,mW (B) \\
\cite{Rana16}    & NI PXI-7833R, xc2v3000 & GL, short memory & $0.2$, $-0.6$, $0.99$; $L$=100, 400 & 2852 slices (19\%), 3854 LUT (13\%), 2538 FF, 25 MULT18$\times$18, 14 BRAM; no $f_{\max}$ & --- \\
\cite{Sharma19}  & FPGA, Virtex-7 VC709 & lattice-wave IIR, ALO-fitted & $r$=0.5; $N$=5 & 882--1097 LUT by encoding; LUT-only design, no slice, DSP or $f_{\max}$ & --- \\
\cite{Malik20, MalikCorrigendum21} & Altera DE2-115 & Al-Alaoui + CFE, direct form II, fractional HR & $0.80$; $n$=1--5 & 3825--9540 logic elements, 3455--8636 reg, 12 mult, 0 memory bits & 317--419\,mW over $n$=1--5 \\
\cite{Tolba19b}  & FPGA, Artix-7 XC7A100T & GL; fixed window, and a linear fit to the tail coefficients & $\pm0.3$--$0.9$; $L$=32, 512, 1024 & FOPID application only: 1136 slices, 3877 reg, 75 DSP, 20.4\,MHz. None for the operator itself & --- \\
\cite{Tolba19c}  & FPGA, Virtex-5 XC5VLX30T & GL + non-standard finite difference, Izhikevich & $(\alpha,\beta)$ to $(0.85,0.85)$; full history & 1245--1734 slices, 0 DSP, 0 RAM, 33.4--36.4\,ns delay, 20-bit & --- \\
\cite{Sacu23}    & TI F28335 DSP at 150\,MHz, two external 12-bit DACs & Laplace--Adomian decomposition, 4 terms, \emph{no history retained} & $q$=0.85, 0.95; no window & no resource figures reported & --- \\
\cite{ValenciaPonce21} & UMC 180\,nm CMOS, \emph{simulated} & 3-pole/2-zero rational, OTA-C cascade & $0.9$; 3 sections & 6 OTA per integrator; no transistor count or area & 149.8\,\textmu W per integrator (sim.) \\
\cite{Abulencia15} & 0.35\,\textmu m CMOS, \emph{layout only} & Valsa RC ladder CPE & $0.25$, $0.50$; $m$=5 & 17 R, 11 C, 26 transmission gates, 12.56\,mm$^2$ layout & --- \\
\hline
\end{tabular}
\label{tab:hardware}
\end{table}


\section{Devices with intrinsic fractional dynamics}
\label{sec:devices}     

\subsection{Where constant phase comes from}

A constant-phase element is a two-terminal device whose impedance
follows $Z(s)\propto s^{-\alpha}$, so that its phase angle stays at
$-\alpha\pi/2$ across a band rather than approaching $-\pi/2$ as an
ideal capacitor's would. Westerlund and Ekstam's point bears repeating
in this context: to first approximation every real capacitor behaves
this way, and the ideal element is the idealization
\cite{Westerlund94}.

Several physical accounts exist and they do not agree. Surface fractality
was an early candidate \cite{Nyikos85}. Brug and colleagues supplied the
conversion between a constant-phase parameter and an effective
capacitance that any comparison across devices requires \cite{Brug84},
and without which the reported values in table~\ref{tab:devices} cannot
be placed on a common footing. Jorcin and coworkers attribute the
behavior to a distribution of local time constants across the electrode
\cite{Jorcin06}. Hirschorn and colleagues developed the account most
useful to a device engineer, in which $\alpha$ follows from the
resistivity profile through a film \cite{Hirschorn10a, Hirschorn10b};
this is the only constant-phase theory in the set that predicts the
order from something a fabricator controls. Lasia's assessment is that
no single origin is universal \cite{Lasia22}, and the distributed RC
picture predates all of it \cite{Nathan73}.

\subsection{Why power-law relaxation is generic in solids}
\label{sec:generic}

The strongest argument that this route is practical rather than exotic
comes from dielectric physics. Jonscher's universal dielectric response
holds that power-law behavior in the frequency domain is the normal case
across an enormous range of solids, not a property of specially prepared
materials \cite{Jonscher77}. Nigmatullin connected that observation to
fractional operators \cite{Nigmatullin84}. Scher and Montroll's
continuous-time random walk account of dispersive transport in amorphous
semiconductors \cite{Scher75} does something more specific and more
useful: it turns the statement that an amorphous oxide device should show
power-law relaxation into a physical prediction with a mechanism behind
it, rather than an observation to be curve-fitted after the fact.
Related treatments are available \cite{Sibatov20, Metzler00}.

For the embodiment strategy this is the enabling fact. Power-law
relaxation is not something a fabricator must engineer against the grain
of the material. It is closer to the default in disordered solids, and
the engineering problem is placing it, not creating it.

\subsection{Fabricated fractional-order capacitors}
\label{sec:fabricated}

Table~\ref{tab:devices} collects the fabricated devices with published
characterization. The materials are diverse and the results are more
consistent than the diversity suggests: reduced graphene oxide in
polymer \cite{Elshurafa13}, molybdenum disulfide in ferroelectric
polymer \cite{Agambayev18}, ferroelectric polymer blends without a
filler \cite{Agambayev17a, Agambayev17b}, multiwall carbon nanotubes in
epoxy \cite{John17}, and copper foil on a polymer nanocomposite
\cite{Shah23}. Other fabrication routes and characterizations are
reported in \cite{Krishna11, Buscarino18, BiswasCB18, Kartci18,
Kartci19}. The passive RC ladder remains the baseline any device has to
beat, and it is a more demanding baseline than it first appears
\cite{Valsa13}.

Three observations follow from reading these papers against one another
rather than in isolation, and they are recorded in the table's structure.

First, every order range in the literature is an envelope across
fabricated variants, not a property of one device. A paper reporting
$\alpha$ from $0.33$ to $0.74$ has usually made five capacitors at five
filler loadings, each holding one near-constant order across the band
\cite{Elshurafa13}. Quoting the range as a single row implies a tunable
device, and no such device has been built. Tuning happens at the
fabrication step.

Second, the constant-phase band and the instrument's sweep window
frequently coincide. Four of the six entries in table~\ref{tab:devices}
report a band whose endpoints are the impedance analyzer's limits, which
means the quoted band is a lower bound on the device and tells us
nothing about behavior outside it. This is not a criticism of the
individual papers, which characterize what their equipment reaches. It
becomes a problem only when the collected values are read as a map of
what materials can do.

Third, the tolerance criterion defining the band is often unstated.
Agambayev and colleagues define bandwidth explicitly as the region where
phase varies by no more than four degrees \cite{Agambayev18}, so their
band and their ripple figure are one claim. Others report a band with no
tolerance attached \cite{Agambayev17a}, which makes the number
incomparable with the first.

That criterion carries a consequence the device literature has not
drawn. Because every order in table~\ref{tab:devices} follows
$\alpha=|\phi|/90^{\circ}$, a stated in-band ripple of $\pm4^{\circ}$ is
an order excursion of $\pm0.044$, which is nine percent of $\alpha$ at
$0.5$ and thirteen percent at $0.33$. Shah and colleagues' $\pm3.1^{\circ}$
to $\pm4.6^{\circ}$ reaches $\pm0.051$ \cite{Shah23}. Set that against the
ten percent order tolerance that section~\ref{sec:gaps} takes from
\cite{Mastin26}. At the low-order end the reported ripple alone meets or
exceeds that tolerance, and at the high-order end it still consumes
about half of it, in both cases before any unit-to-unit spread or
long-term drift has been measured. The problem is not only that
variability goes unpublished. Part of the budget is already spent by the
numbers that are published.

The emulator route sidesteps the materials problem entirely. Tsirimokou
and colleagues fabricated and measured CMOS constant-phase emulators at
five orders from $0.3$ to $0.7$, verified from $10$ to $300$\,Hz
\cite{Tsirimokou16}, and further CMOS realizations have followed
\cite{SciRep25MOS}. These reach exactly the corner of order and
frequency that the passive devices do not. The caveat is fundamental
rather than incidental: an emulator is an active circuit that supplies
the order on demand while consuming static power and area, which
forfeits the argument for embodiment completely. It computes the kernel
rather than being it. Notably, the letter reports no power figure at
all, which for an active part is the number that decides whether the
trade is worth making. Fouda and coworkers offer a more interesting
hybrid, deriving a constant-phase element from a memristive crossbar
\cite{Fouda21}, which would place the fractional operator in the same
substrate as the neuromorphic array. Foundational circuit treatments are
in \cite{Elwakil10, RadwanSalama12, Radwan12, Tavazoei20}.

\begin{table}
\caption{Fabricated constant-phase devices and one CMOS emulator. In every entry
the order range is an envelope across several fabricated variants at different
filler loadings or polymer blends, never one device tuned across frequency; the
number of variants is given so the envelope is not misread as tunability. Orders
follow each paper's own conversion $\alpha=|\phi|/90^{\circ}$. An em-rule marks a
quantity the source does not report. Note that four of the six bands coincide
with the impedance analyzer's sweep window, so they bound the measurement and not
the device.}
\centering\footnotesize
\begin{tabular}{l p{2.5cm} p{1.5cm} p{2.4cm} p{1.4cm} p{2.4cm}}
\hline
Work & Material (variants) & $\alpha$ & Band (Hz) & Ripple & Drift \\
\hline
\cite{Elshurafa13} & rGO--polymer (5 loadings) & 0.33--0.74 & $5\times10^{4}$--$2\times10^{6}$, continuous & --- & --- \\
\cite{Agambayev18} & MoS$_2$--ferroelectric polymer (3) & 0.64--0.90 & $10^{2}$--$10^{7}$, continuous & $\pm4^{\circ}$, and the band's definition & --- \\
\cite{John17} & MWCNT--epoxy (5 loadings) & 0.50--0.94 & \textbf{three disjoint}: 110--1100; $10^{4}$--$1.18\times10^{5}$; $2.3\times10^{5}$--$2\times10^{7}$ & $\pm0.9^{\circ}$--$\pm4.2^{\circ}$ & \textbf{$\pm2^{\circ}$ over 6 months} \\
\cite{John17}, 60\,$^{\circ}$C cure & MWCNT--epoxy (1) & 0.593 & 550--$2\times10^{6}$, continuous & $\pm1.5^{\circ}$ & as above \\
\cite{Shah23} & Cu foil on PVDF--GNS/rGO (6) & 0.61--0.88 & per device; best $1.3\times10^{3}$--$10^{7}$ & $\pm3.1^{\circ}$--$\pm4.6^{\circ}$ & unit-to-unit $\pm10\%$ on $C_{\alpha}$; long-term --- \\
\cite{Agambayev17a} & PVDF, P(VDF-TrFE), terpolymer and blends (12) & 0.69--0.90 & $10^{5}$--$10^{7}$, continuous & --- (no tolerance defined) & --- \\
\cite{Tsirimokou16} & CMOS emulator, AMS 0.35\,\textmu m (5 fixed orders) & 0.3--0.7 & 10--300, continuous & $\le5^{\circ}$ error & --- \\
\hline
\end{tabular}
\label{tab:devices}
\end{table}

\subsection{Supercapacitors and double-layer devices}
\label{sec:edlc}

One term needs unpacking before this subsection can do its work, because
section~\ref{sec:operators} assumed no fractional calculus and it would
be inconsistent to assume electrochemistry here. Bring an electrode into
contact with an electrolyte and the ions in the liquid arrange themselves
against the electronic charge in the solid, over a region a few
nanometres thick. That region is the \emph{electrochemical double layer},
and it stores charge electrostatically rather than through a redox
reaction. A device built on it is an electric double-layer capacitor, or
in commercial language a supercapacitor.

Why such a device is fractional follows from the same argument as
figure~\ref{fig:mechanisms}. The electrode is porous, so ions reaching
different pores travel different distances through different geometries,
and the interface presents not one $RC$ time constant but a broad
distribution of them. That distribution is what produces constant phase,
and electrochemists have characterized these devices in exactly those
terms for years \cite{Martynyuk15, Freeborn13a, Allagui16, Allagui18,
Allagui20}, to the point of a dedicated review of their fractional-order
electrical characterization \cite{Allagui18}. They matter here for one reason that the thin-film literature does not
share, and the reason is mechanistic rather than incidental. Ion
transport through a porous electrode is slow, so the time constants that
distribution spans are long, and the constant-phase behavior therefore
sits at low frequency: precisely where neuromorphic adaptation lives, and
where, as figure~\ref{fig:mismatch} shows, the thin-film devices do not
reach. A thin dielectric film has the opposite problem. Its relaxation
comes from charge redistribution over nanometres rather than ion motion
through micrometres, so its time constants are short and its band sits
high.

This family is cited here but not entered in
table~\ref{tab:devices}, and the reason bears on how the rest of this
section should be read. Electrochemical impedance spectroscopy on
double-layer electrodes routinely runs into the millihertz range, so the
low-frequency corner of figure~\ref{fig:mismatch} is not unexplored
territory in any absolute sense. It is unexplored by the thin-film and
solid-state devices that a neuromorphic process could integrate. We have
not tabulated the double-layer measurements because they are reported to
conventions that do not map cleanly onto the order-and-band columns used
here, and assembling that mapping is a piece of work in its own right.
Section~\ref{sec:gaps} lists it as one.

That is not an incidental advantage. It is why the only end-to-end
demonstration in this field, discussed below, was built from
supercapacitor stacks rather than from any of the thin-film devices in
table~\ref{tab:devices} \cite{VazquezGuerrero24}. The cost is size:
these are discrete components measured in millifarads, not integrable
elements, and nothing about the double-layer mechanism suggests it will
shrink onto a die. The low-frequency corner is currently reachable only
with parts that cannot be integrated.

\subsection{Memristors, memcapacitors and ionic devices}

Devices developed for neuromorphic memory offer a second route, and one
already compatible with the substrate. Diffusive memristors relax
through ion migration after a stimulus is removed \cite{WangZ17}, and
the mechanism is the same one-dimensional diffusion picture that appears
in section~\ref{sec:biology} as a candidate explanation for channel
gating statistics \cite{Millhauser88}. Chang and colleagues report a
stretched-exponential transition from short-term to long-term memory in
a tungsten oxide device \cite{Chang11}, which is the closest thing in
the memristor literature to a device computing a memory kernel rather
than storing a weight. Related short-term dynamics are reported in
\cite{Ohno11, KimS15}. Memcapacitive devices have recently been reviewed
for neuromorphic use \cite{AbuHamra25, ChenKT25}, and electrochemical
random-access memory and other ionic devices supply a further family
with intrinsically slow dynamics \cite{Fuller17, Onen22, Tang18,
vandeBurgt17}. One caution on attribution: the fractional memristor
model of \cite{Fouda15} is a model, not a fabricated device, and is
sometimes cited as though it were the latter.

\begin{figure}
  \centering
  \includegraphics[width=\textwidth]{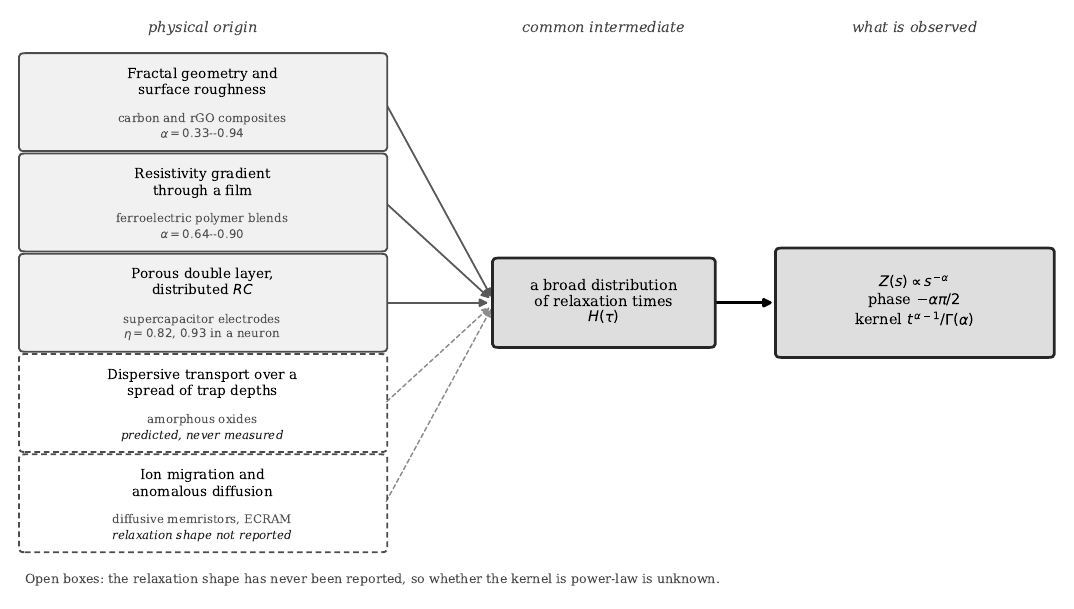}
  \caption{Physical routes to a power-law kernel. Several unrelated processes
  reach the same place: each produces a broad distribution of relaxation times,
  and it is that distribution, rather than any particular microscopic
  mechanism, which yields constant phase and a power-law memory kernel. This is
  why section~\ref{sec:generic} can claim the behavior is close to generic in
  disordered solids rather than a property of specially prepared materials.
  Solid boxes mark routes whose orders have been measured, with the range
  observed across the devices in table~\ref{tab:devices}. Open boxes mark the
  two routes that have not: dispersive transport predicts power-law relaxation
  in amorphous oxides but the prediction has not been tested against a
  fabricated fractional element, and the ionic and memristive devices that the
  neuromorphic community already builds are characterized for retention,
  endurance and switching energy rather than for the shape of their relaxation.
  Closing that second gap is the third item of section~\ref{sec:outlook}.}
  \label{fig:mechanisms}
\end{figure}

The gap in this literature is specific and worth naming. These devices
are characterized for retention, endurance and switching energy, which
are the figures of merit for memory. Almost none are characterized for
the shape of their relaxation, which is the figure of merit here. Whether
a given mechanism yields power-law or stretched-exponential or simply
exponential decay is usually not reported, and the distinction decides
whether the device embodies a fractional kernel or merely a slow one.
Establishing that mapping across the mechanisms in this subsection would
be a substantial contribution and nobody has made it.
Figure~\ref{fig:mechanisms} places these devices against the routes whose
orders have been measured, and the contrast is the point: the two families
closest to a CMOS process are the two whose relaxation shape is unknown.

\subsection{The order and bandwidth mismatch}
\label{sec:mismatch}

One study closes the loop from a physical fractional element to a neuron
circuit to a measured neuronal behavior. Vazquez-Guerrero and colleagues
built fractional integrate-and-fire and Hodgkin--Huxley gate circuits
using series stacks of supercapacitor elements, measured operating
orders of $\eta=0.82$ for the integrate-and-fire membrane and
$\eta=0.93$ for the potassium gate, and reproduced power-law
spike-timing adaptation, history-dependent rate accommodation and
spectral whitening, validated against recordings from weakly electric
fish \cite{VazquezGuerrero24}. It is the existence proof this field has
been missing, and everything below should be read as a question about
how to generalize it rather than whether the approach can work at all.

Assessing the remaining distance requires separating two targets that
the literature routinely conflates.

The first is the \emph{descriptive} order: the value that characterizes
a biological neuron, near $0.15$ for cortical pyramidal cells
\cite{Lundstrom08}, with the companion models of
\cite{VazquezGuerrero24} calling for $0.2$ to $0.3$. The second is the
\emph{task-optimal} order: the value that maximizes performance on a
workload. These are different quantities and there is no reason they
should agree. Sweeping the order across three benchmarks in a
fractional-order reservoir, Mastin and colleagues find the optimum near
$0.3$ to $0.5$ for spoken-digit classification, with offline and online
readouts both peaking at $0.5$ so the result is not an artifact of the
training procedure, near $0.5$ to $0.7$ for cart-pole control, and near
$0.1$ for a physiological prediction task whose accuracy falls
monotonically with the order \cite{Mastin26}.

That result is the present author's own, so it should not be asked to
carry this argument by itself, and as of this year it does not have to.
Olivares and Santamaria have swept the order independently, on the harder
variant of the same cart-pole benchmark in which the controller sees only
position and angle, at seven orders from $0.2$ to $1.0$ with forty
evolutionary runs at each \cite{Olivares26}. Their optimum falls at
$\alpha=0.5$, with usable performance from $0.35$ to $0.80$, and their
learning rate peaks at the same value.

Two groups, different neuron models, different training procedures, the
same benchmark, the same answer. That is the strongest evidence in this
review for any quantitative claim it makes, and it is worth saying what
makes it strong: neither group set out to confirm the other, and the
agreement is on a number rather than on a direction. The two share no
personnel and no analysis, though both efforts draw on the same National
Science Foundation grant, which a reader weighing that independence
should know.

The convergence is more informative than the optimum alone, because both
sweeps are non-monotonic in the same way. Olivares and Santamaria report
that only $8$ of $40$ runs succeeded at $\alpha=0.95$ and $5$ of $40$ at
$\alpha=0.20$, against $31$ of $40$ for the integer-order network, so the
advantage is confined to a middle band and vanishes at both ends. Their
reading is that at high order the memory fades too fast to help, and at
low order it swamps the present input. An engineer reading only that
fractional order improves performance would take the wrong lesson. The
order has to be in the band, and the band is now bounded from two
directions by two groups.

A third group reports the same shape of result from a different
direction. Ge and colleagues tune the order per dataset across six graph
benchmarks and select $0.3$ on two of them, $0.8$ on two more and $0.9$
on a fifth \cite{Ge26}. The two remaining datasets select $\alpha=1$,
which by their own remark is the integer-order limit of their scheme, so
on a third of their benchmarks the tuned optimum is to switch the
fractional dynamics off. That is the more useful half of the table. It is
a tuned selection rather than a controlled sweep and carries less weight
than the two studies above, but three groups now place the useful orders
in a band running from about $0.3$ to $0.9$, and one of them reports the
cases where no such band exists.

\begin{figure}
  \centering
  \includegraphics[width=0.85\textwidth]{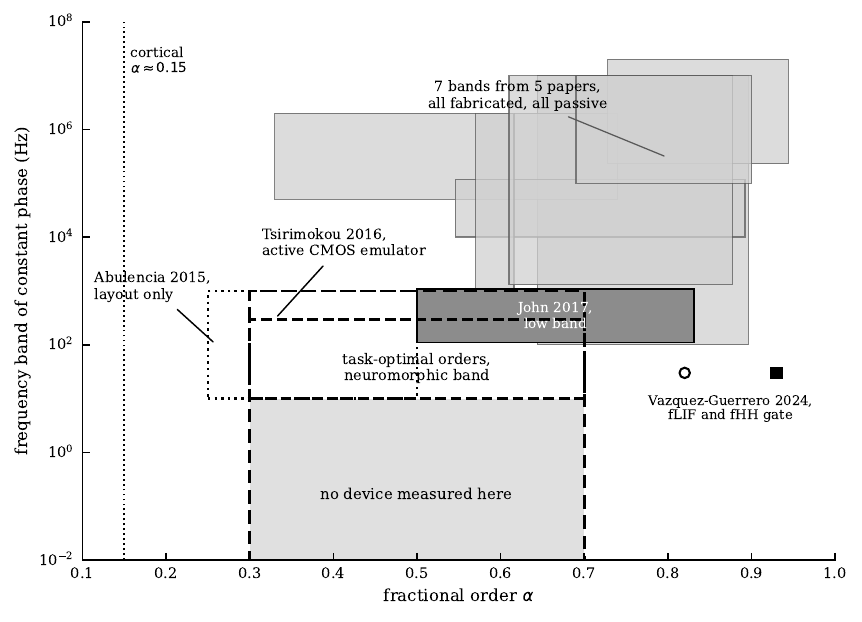}
  \caption{Fractional order and frequency band of constant-phase behavior
  for every device surveyed in table~\ref{tab:devices}, against the dashed
  region a neuromorphic design would need: the task-optimal orders reported
  by \cite{Mastin26}, across the band adaptation occupies from milliseconds
  to tens of seconds. The vertical extent of that region is an inference
  rather than a measurement: the order sweeps of \cite{Mastin26} are run
  in discrete time at $\Delta t=0.1$ over $25$ macro-steps per sample, and
  are placed here on the assumption that a macro-step corresponds to a
  millisecond of biological time, which is the convention their
  spike-timing comparisons use. A different assumed timestep slides the
  region vertically. The band this region occupies is set by the sensory
  input rather than by the simulation timestep, for the reason
  section~\ref{sec:mismatch} gives at its close. Each box is an envelope over the fabricated variants
  of one paper, not one tunable device, and orders follow each paper's own
  conversion $\alpha=|\phi|/90^{\circ}$. The order axis is no longer the
  obstacle: fabricated passive devices span it. They sit above the band,
  and the only parts reaching into it are an active CMOS emulator that
  reports no power figure, one unfabricated layout, and the lowest of
  \cite{John17}'s three disjoint bands. Below roughly $10$\,Hz nothing has
  been measured at all, which is where the slowest adaptation lives. The
  two marked points are the operating orders of the fractional
  integrate-and-fire and Hodgkin--Huxley gate circuits of
  \cite{VazquezGuerrero24}, whose companion models require
  $\eta\approx0.2$--$0.3$.}
  \label{fig:mismatch}
\end{figure}

One caveat survives for both. These are simulation results, run in
discrete time where the operative parameter is retained history in
samples rather than a physical bandwidth, so mapping an optimal simulated
order onto a requirement for a fabricated element involves an assumed
timestep, stated in the caption of figure~\ref{fig:mismatch}.

One distinction should be carried forward deliberately.
Section~\ref{sec:gaps} criticizes how \cite{Mastin26} matched its
integer-order baseline, and that criticism is sound. It bears on the
fractional-versus-integer comparison. It does not bear on where the
optimum sits, because the sweep is internally controlled: one
architecture, one training budget, seven orders. A reader should not
carry the objection across, and we would rather say so than let the
paper's best quantitative claim dissolve by association.

So is the order the problem? Against the second target it is not.
Fabricated devices span $\alpha$ from $0.33$ to $0.94$, which covers the
optima of two of those three benchmarks with room on either side. The
order gap has closed. We state that plainly because it is the opposite
of the field's working assumption, and because believing otherwise sends
effort to the wrong axis.

What survives is narrower and, we think, more useful. The binding
constraint is the frequency band. Neuromorphic adaptation runs from
milliseconds to tens of seconds, which is to say from roughly a
kilohertz down towards hundredths of a hertz, while the thin-film
devices in table~\ref{tab:devices} hold their constant phase from about
$100$\,Hz upward. Figure~\ref{fig:mismatch} shows the position of every
tabulated device against the region a neuromorphic design would need.
The sub-kilohertz region is not empty: \cite{Agambayev18} reaches
$100$\,Hz, the lowest of \cite{John17}'s three disjoint bands runs from
$110$ to $1100$\,Hz, and both CMOS entries operate below a kilohertz.
Below about $10$\,Hz, no tabulated device has been characterized.

It matters what that does and does not say, and we want to be exact
because the looser version of this claim is tempting and wrong. The
low-frequency corner is \emph{not} unoccupied. Electrochemical
double-layer electrodes work there, their impedance is routinely measured
into the millihertz range, and section~\ref{sec:edlc} is the reason the
only end-to-end neuron demonstration in this field was built from
supercapacitor stacks \cite{VazquezGuerrero24}. What has not been
characterized there is any device a neuromorphic process could
\emph{integrate}. Every entry in table~\ref{tab:devices} is either a
thin-film element measured no lower than about $100$\,Hz, or an active
CMOS part, and the one family that reaches the corner does so as discrete
millifarad components.

So the gap is not an absence of measurement in general. It is the absence
of an integrable device that has been measured there, and it is worth
asking why. Four of six tabulated papers report a band coincident with
their instrument's sweep window, so in those cases the lower endpoint
describes the measurement and not the material. Low-frequency impedance
spectroscopy is standard in the adjacent electrochemical literature. What
nobody appears to have done is to point that instrument at a thin-film
fractional-order capacitor and report the result.


One assumption underneath all of this deserves to be stated rather than
smuggled. Nothing in physics requires a neuromorphic circuit to run on
biological wall-clock time. Clock it a thousand times faster than cortex
and the dashed region of figure~\ref{fig:mismatch} slides up three
decades, into the band the fabricated devices already occupy, and the
mismatch this section has spent several pages establishing disappears. So
why not do that?

Because the timescale is not set by the circuit. It is set by whatever
the circuit is listening to. A system that adapts to the statistics of an
event camera, a microphone, a tactile array or an electrode does so
against an input whose spectrum is fixed by the physical world, and the
world cannot be clocked faster. Adaptation that normalizes a stimulus
statistic has to operate over the interval on which that statistic
actually varies, which at the slow end of sensory adaptation is seconds
to tens of seconds, whatever the substrate is doing internally.

One could buffer the input and replay it compressed. That is precisely
the move the embodiment argument exists to avoid.
Section~\ref{sec:cost} derived the cost of fractional dynamics as the
storage of retained history, and the whole case for a physical
constant-phase element is that the device holds that history in its own
relaxation rather than in registers. Time-compressing the input
reinstates the storage, and with it the cost the device was meant to
remove. The faster clock buys nothing it does not immediately pay back.

The requirement is real, then, but it is not universal, and the scope is
worth stating plainly. It binds for hardware that interfaces to the
physical world in real time, which is the setting the embodiment argument
assumes and the setting in which a passive fractional element earns its
place. It does not bind for a system whose inputs are synthetic or
replayed from storage. For that case the bandwidth argument of this
section does not apply, and the order axis is the only one that matters.

Two further obstacles are real and should not be minimized. A residual
order gap remains for long-memory tasks whose optimum sits near $0.1$
and for matching the descriptive biological order, and no fabricated
passive device reaches $0.15$ to $0.30$ with usable bandwidth in the
neuromorphic range. And almost nothing in table~\ref{tab:devices} is
compatible with a CMOS process. The two CMOS-native entries are an
emulator and an unfabricated layout, so the only route that integrates
today is the one that abandons the embodiment argument.

\subsection{What the embodiment argument actually buys}
\label{sec:embodiment}

It is worth restating the case for devices in its corrected form,
because the strong version of the claim does not survive the preceding
sections and the weaker version is still substantial.

What disappears is the runtime cost of carrying history, and the honest
comparison is no longer against the direct form.
Section~\ref{sec:strategies} grants the fast schemes their
$\mathcal{O}(\log N)$ accumulators, so a device is not escaping a buffer
of $N$ samples; it is escaping a handful of accumulators, the arithmetic
that updates them, and the clock that steps it. Put in storage terms the
saving is modest. What it buys is different in kind: the relaxation is
the state, so there is no update to schedule, no word length to choose,
and no coefficient to represent. The precision wall of
section~\ref{sec:precision} does not shrink for a device, it stops
applying, because there are no coefficients to underflow. That is the
version of the claim that survives the preceding sections, and it is the
reason the device route deserves the attention this review gives it.

What is bought is not a free kernel but a cheap and fixed one. The
runtime cost becomes a commitment made at fabrication to whatever order
and band the process yields. Section~\ref{sec:digital} describes designs
in which the order is a parameter that can be changed by rewriting a
coefficient table; here it is a material property. The device state
cannot be read out, reset, or checkpointed, which rules out the
initialization control that section~\ref{sec:operators} identified as
already delicate for fractional systems. And the drift and
unit-to-unit variability of that fixed order are, with one partial
exception in table~\ref{tab:devices}, unmeasured.

There is now a measured version of that last point, and it is more
concrete than an argument from the operator alone. Olivares and
Santamaria found their fractional controllers failing early in an episode,
before the memory trace had filled, at a rate that rose as the order fell
\cite{Olivares26}. Their fix was to hand control to an integer-order
network for the first six decisions and let the fractional network watch,
then take over, which recovered as much as thirty-five seconds of balance
time at $\alpha=0.5$. A fractional controller, in other words, needed a
conventional one to nurse it through the first sixty milliseconds of every
trial.

Section~\ref{sec:operators} noted that the operator depends on its lower
terminal, so there is no fractional analog of clearing a state variable.
This is what that costs in practice. A device whose memory lives in a
physical relaxation cannot be initialized, and a system built from such
devices pays a warm-up penalty on every restart that the literature has
not previously quantified.

Framed as a trade rather than a triumph, the obstacles turn into
targets: (1) bring the order down to about $0.2$ while keeping usable
bandwidth; (2) move the band into the range from hundredths of a hertz
to a hundred hertz; and (3) publish drift and unit-to-unit variability
data. Each is a concrete experiment rather than a research direction,
and section~\ref{sec:gaps} argues that the third is both the cheapest
and the most overdue.

\section{What is the field not measuring?}
\label{sec:gaps}                      

\subsection{Learning, and a null result}

Fractional gradient descent is scoped out of this review. The method is
contested at the level of basic mathematics: replacing the gradient with
a fractional derivative moves the fixed point, so minimizing
$f(u)=(u-c)^2$ by plain fractional gradient descent converges to
$c(2-\alpha)$ rather than to $c$ \cite{Wei20, Ferreira26}, an early
proposal was refuted in the journal that published it \cite{Pu15,
WahabKhan20}, and the fractional least-mean-squares algorithm received
the same treatment \cite{Bershad17}. The standard repairs, truncating the
memory or the series, remove the non-locality that motivated the method,
which is section~\ref{sec:digital}'s hardware trade reappearing in
software. Careful positive results exist \cite{Shin23} and the variants
have been surveyed \cite{Elnady25, Wang17}, but the area is not settled
enough for a hardware review to report convergence properties.

The null result is the more useful half. No FPGA, ASIC, memristive or
crossbar implementation of any fractional learning rule appears in the
surveyed set, and the most recent fractional spiking network we are aware
of trains offline by evolutionary search on a compute cluster rather than
by any plasticity rule at all \cite{Olivares26}. The fractional hardware literature implements dynamics; it
does not implement learning.

Two papers from 2026 look like counterexamples and repay a closer
reading. Yin and colleagues title theirs fractional-order plasticity, and
what is plastic is the order rather than the rule: a per-neuron order
parameter is trained jointly with the weights by ordinary spatiotemporal
backpropagation \cite{Yin26}. After they fold the order-dependent
coefficient into the weights, which is a step they take themselves to
stabilize the gradients, their membrane update reads
$V_i^l(t)=\hat{\tau}_{ci}^{l}V_i^l(t-1)+\sum_i
s_i^{l-1}(t)\hat{\omega}_{ij}^{l-1}$. That is a leaky integrate-and-fire
neuron with a trainable leak, and it carries no history sum. What their
results establish is the value of heterogeneous learnable time constants,
which is an integer-order finding of some standing and which
section~\ref{sec:gaps} treats as the competition. Ge and colleagues do
carry a genuine power-law kernel, with coefficients decaying
algebraically in the lag, and train it by surrogate-gradient
backpropagation that is integer-order throughout \cite{Ge26}. Neither
runs on hardware. That the two most visible recent papers with
fractional in the title both train by ordinary gradient descent is
better evidence for the gap than our failure to find something. Whether the claimed convergence advantages
survive fixed-point quantization is therefore open by default, not
because the experiment gave an ambiguous answer but because we found no
one who has run it.

\subsection{Reservoir computing is the better place to plant a flag}

Fractional-order reservoir computing is in considerably better shape
than fractional backpropagation, and it exists independently of any one
group \cite{Kobayashi18, Yao20, Yao24, Yao26}. It also fits the physics.
Physical reservoir computing already takes the position that a
substrate's own dynamics do the computation \cite{Tanaka19, Cucchi22},
and the hardware reservoirs built on memristive, spintronic and delay
substrates \cite{Du17, Moon19, Torrejon17, Appeltant11} rely on exactly
the device relaxation that section~\ref{sec:devices} wants to exploit.
A fractional reservoir asks the substrate for a property it may already
have, rather than asking a digital pipeline to synthesize one.

The opening is that all of it is software. The order sweeps that
establish task-optimal values, including those in \cite{Mastin26}, are
simulations. Nobody has built a reservoir whose nodes are physical
fractional elements and measured whether the order-dependence survives
device variability. That is a well-posed experiment with an existing
theoretical prediction to test against, and it is the shortest path from
this literature to a result that would interest someone outside it.

\subsection{Gaps, stated as questions}

Each of the following is answerable with existing methods, and we state
them as experiments rather than as complaints.

First, there is no atlas of the fractional order in neurons. The value
near $0.15$ comes from a small number of cell types in one preparation.
How does it vary across cell class, cortical layer, brain region and
species? Second, the two fractional Hodgkin--Huxley formalisms, with the
order on the gates \cite{Teka16} and on the capacitance
\cite{Weinberg15}, have never been fitted to the same recordings. Which
placement better predicts held-out data? Third, of the fabricated constant-phase devices tabulated here, only one
publishes long-term stability data \cite{John17} and one a capacitance
spread \cite{Shah23}; none publishes unit-to-unit variability of the
order itself. This is the cheapest gap on the list
and the tolerance target is already known: Mastin and colleagues report
under two percentage points of accuracy loss at ten percent order drift,
so a device holding its order to that tolerance is good enough, and
nobody has checked whether existing devices do.

Fourth, no work in the surveyed set reports energy per fractional operation. This is
not a minor omission. It makes structured comparison against
integer-order neuromorphic hardware \cite{Davies18, Merolla14} or
against analog in-memory computing impossible in principle rather than
merely inconvenient, and any claim that fractional order is efficient is
currently unfalsifiable. Fifth, no study sweeps approximation order
against resource count and accuracy on a common platform, so the
central engineering claim of the substitution strategy is unquantified.
Sixth, there is no benchmark task and no real workload; error is
routinely reported against a software simulation of the same equations,
which measures numerical fidelity and not usefulness. Seventh,
fractional spike-timing-dependent plasticity does not exist as a body of
work, which given the memory-dependence of the underlying learning rule
is a conspicuous absence.

One further gap follows from what this review could not settle. Nobody
has tabulated the double-layer literature
\cite{Martynyuk15, Freeborn13a, Allagui16, Allagui18, Allagui20} in the
order-and-band form of table~\ref{tab:devices}. Those electrodes occupy
the low-frequency corner that every integrable device misses, so the
translation is not bookkeeping. It would establish what order is
achievable down there at all, which is the number a materials programme
aimed at an integrable equivalent would need as its target.

A further gap has just been created rather than closed. Now that two
groups agree on the task-optimal order for cart-pole
\cite{Mastin26, Olivares26}, the obvious question is whether the
agreement survives a second task. Both sweeps that reach $\alpha=0.5$ are
closed-loop control; the one benchmark where the reported optimum sits
elsewhere, near $\alpha=0.1$, is a physiological prediction task with long
correlations \cite{Mastin26}. Whether the optimum is a property of the
task's memory structure or of control problems in particular is now the
sharpest open question in the software half of this field, and it is
answerable with a laptop.

One more is a precondition for everything in section~\ref{sec:devices},
and we should have stated it before the table rather than after. Every
order in table~\ref{tab:devices} comes from small-signal impedance
spectroscopy, which is to say from a low-amplitude sinusoid swept in
frequency. A spiking circuit drives the element with large, step-like,
non-sinusoidal transients, and double-layer devices in particular are
bias- and rate-dependent. Whether an $\alpha$ extracted from an impedance
sweep predicts the time-domain kernel under spiking excitation is
untested for every entry in that table. V\'azquez-Guerrero and
colleagues come closest and only for supercapacitor stacks
\cite{VazquezGuerrero24}; for thin films there is nothing. The
experiment is a step response and a fit, on devices that already
exist.

\subsection{A metric set and two benchmark tasks}

We propose three measurements that would make results comparable.
Energy per fractional state update, reported with the order, the
retained history and the error tolerance it was measured at, since a
figure without those three is not interpretable. Effective order and its
ripple over an explicitly stated band, with the tolerance that defines
the band, which would resolve the incomparability documented in
table~\ref{tab:devices}. And accuracy against an integer-order baseline
of \emph{matched parameter count} rather than matched architecture.

That last point carries the most weight and is the easiest to get wrong,
and there is now a concrete case to point at. Olivares and Santamaria
report their fractional network as fifty percent smaller than the
integer-order baseline it outperforms, and they are open about how the
comparison was set up: the baseline needed twelve hidden neurons against
the fractional network's six, giving thirty-seven free parameters against
nineteen, and it was trained for six hundred evolutionary generations
against one hundred and fifty, with a different membrane time constant
and different mutation and weight ranges chosen, in their words, to
improve the baseline's success \cite{Olivares26}. Every one of those
choices is defensible on its own terms and every one is reported, which
is more than most papers manage. But the resulting comparison is between
networks matched on neither parameter count nor training budget, so it
cannot settle whether the advantage comes from the fractional operator or
from the search. Neither the six-neuron integer-order control nor the
twelve-neuron fractional one is reported.

We raise it because it is the best-documented instance of the problem
rather than the worst, which is rather the point. The honest competitor
for a fractional neuron is not a plain leaky integrate-and-fire unit. It
is an adaptive one, or a heterogeneous population, given the same number
of free parameters and the same optimization effort. Comparisons that
miss that flatter the fractional approach for reasons which have nothing
to do with fractional calculus.

For tasks, we suggest spoken-digit classification and cart-pole control,
both used in \cite{Mastin26}, on the grounds that they span
classification and closed-loop control, have published order sweeps to
compare against, and are small enough to run on hardware that exists.

\subsection{The competition this has to beat}

The problem fractional order claims to solve, computation over many
timescales, is being solved without fractional calculus. Perez-Nieves
and colleagues show that simply distributing membrane time constants
across a population improves learning and robustness
\cite{PerezNieves21}. Adaptive spiking neurons achieve long temporal
memory and competitive accuracy on demanding sequence tasks
\cite{Bellec18, Yin21}. Long short-term memory has been implemented on
neuromorphic silicon directly \cite{Rao22}. In each case the mechanism
is integer-order heterogeneity, and in each case the hardware exists
today on platforms with mature toolchains \cite{Davies18, Davies21,
Merolla14, Modha23, Furber14, Pehle22, Pei19, Moradi18}.

The fractional case therefore cannot rest on capability, since the
capability is available by other means. It has to rest on cost: whether
a single physical element supplying many relaxation timescales is
cheaper in area or energy than the explicit heterogeneity that achieves
the same effect. That is an empirical question, it is the question the
missing energy measurements would answer, and at present the field
cannot answer it. Broader context on the neuromorphic landscape is
available in \cite{Christensen22, Kudithipudi25, Schuman22, Mehonic22,
Indiveri15} and on in-memory computing in \cite{Sebastian20, Ielmini18,
Yao20b, Wan22, WangZ18}.

\section{A path forward and a call for action}
\label{sec:outlook}

Where does that leave the field? Three tracks follow from the preceding
sections, and we state each as a target rather than a direction, because
directions are what this area already has plenty of.

\emph{a) Take the electrochemists' instrument to a thin-film device.}
Characterize the fabricated constant-phase elements of
table~\ref{tab:devices} below $10$\,Hz, down to the millihertz range that
impedance spectroscopy on double-layer electrodes reaches as a matter of
routine. This is a measurement campaign and not a fabrication campaign,
it needs equipment that many electrochemistry groups already own, and it
would settle whether the lower edge of every band in
table~\ref{tab:devices} is a property of the material or of the
instrument that was pointed at it. The same visit should answer the question
section~\ref{sec:gaps} raises last. Every order in
table~\ref{tab:devices} comes from a swept small-signal sinusoid, while
a neuron circuit drives its elements with large step-like transients, so
a step response and a fit on the same samples would establish whether
the impedance order predicts the time-domain kernel at all. That is an
afternoon once the device is on the bench. We regard this as the
cheapest and most overdue experiment in the field, and note that the two communities
who would have to cooperate on it are among those
section~\ref{sec:intro} identifies as not reading each other.

\emph{b) Publish drift and variability.} No fabricated fractional-order
capacitor has published unit-to-unit variability of its order, and only
one has published long-term stability data at all \cite{John17}. The
tolerance target is already known. Mastin \etal{} report under two
percentage points of accuracy loss at ten percent drift in the realized
order \cite{Mastin26}, so a device holding its order to that tolerance is
good enough. Whether existing devices do is unknown, and
section~\ref{sec:fabricated} gives a reason to expect the answer to be
tight: a $\pm4^{\circ}$ in-band ripple, which is the criterion these
papers set themselves, is already about half that budget at
$\alpha=0.88$, nine tenths of it at $\alpha=0.5$ and more than all of it
at $\alpha=0.33$, before any unit-to-unit spread or long-term drift has
been measured.

\emph{c) Map relaxation shape onto device mechanism.} Memristive,
memcapacitive and ionic devices are characterized for retention,
endurance and switching energy, which are the figures of merit for
memory rather than for us. Whether a given mechanism yields power-law,
stretched-exponential or plain exponential relaxation decides whether
the device embodies a fractional kernel or merely a slow one, and that
mapping does not exist. It is a table's worth of value and nobody has
built it.

\emph{d) Report energy per fractional operation.} Not one work surveyed
here reports it. Until somebody does, no claim that fractional order is
efficient can be checked, and comparison against integer-order
neuromorphic hardware \cite{Davies18, Merolla14} stays impossible in
principle rather than merely inconvenient.

Two further targets belong to the digital and systems tracks. Sweep the
order against retained history, coefficient word length, resources,
accuracy and energy on a single platform, and report it completely
enough to reproduce, because section~\ref{sec:precision} shows that two
independent limits bound these designs and no published work sweeps them
together. And build a fractional reservoir in hardware. The theory
predicts a task-dependent optimal order, the prediction has been tested
in simulation, and the devices to test it physically already exist. That
is the shortest path to a result someone outside this field would care
about.

So, is fractional-order hardware ready for neuromorphic systems? We
think the answer is not yet, and the reason is not the one the field
expects. Fractional-order dynamics remain a physically grounded route to
multi-timescale memory, and they have one end-to-end experimental
demonstration \cite{VazquezGuerrero24}. The order gap that everybody
worries about has largely closed. What stands in the way instead is a
bandwidth gap of about three decades, an unmeasured low-frequency
corner, and a complete absence of energy data. None of those is a proof
that the approach fails. All three are reasons nobody can yet say
whether it succeeds, and that is a sharper and more actionable position
than the one the field currently occupies.

%


\ack{This review was prepared with the assistance of Claude Opus 5
(Anthropic), used as follows. It produced first drafts of
sections~\ref{sec:biology} to~\ref{sec:outlook}, which the author then
rewrote; extracted the values in tables~\ref{tab:models},
\ref{tab:hardware} and~\ref{tab:devices} from the source articles, with
the page of origin recorded for every entry; performed the numerical work
behind tables~\ref{tab:cost} and~\ref{tab:precision} and
figures~\ref{fig:kernel} to~\ref{fig:mechanisms}, including the
coefficient-precision limit of section~\ref{sec:precision} and the
passivity bound of section~\ref{sec:models}; and checked the bibliography
against the source articles, correcting entries where they disagreed. The
reference list was not generated by the model: every entry originates from
a source document, and no citation was created from the model's own
output. All numerical claims are reproducible from the archived scripts,
and every tabulated value is traceable to a page of a source article. The
author read every source cited for a substantive claim, verified the
derivations, and takes full responsibility for the content of this
review.}


\funding{This material is based on work supported by the National Science
Foundation under grant no.~2318139.}

\roles{C.T. is the sole author and is responsible for
conceptualization, investigation, visualization, and the writing of the
original draft and its revision.}

\section*{Conflicts of interest}

This review cites \cite{Mastin26}, on which
the author is a co-author, and section~\ref{sec:mismatch} depends on the
order sweep reported there. The independent replication of that sweep by
Olivares and Santamaria \cite{Olivares26} was supported in part by
National Science Foundation grant no.~2318139, which also supports the
author's work and this review; the two groups share no personnel and no
analysis for this paper. The author has no financial interest in any device, material or
platform discussed.

\section*{Ethics}

This review is a synthesis of the published literature. It involved no
studies of human participants, human data or tissue, and no animal
studies, so no ethics approval was sought or required. No new data were
collected.

\data{No experimental data were generated. The numerical results in this
review, namely the retained-history bound of table~\ref{tab:cost}, the
coefficient-precision limits of table~\ref{tab:precision}, the
recomputation of the fixed-point table of \cite{Mastin26}, the passivity
calculation of section~\ref{sec:models}, and the data behind
figures~\ref{fig:kernel} to \ref{fig:mechanisms}, were produced by the
short scripts supplied as supplementary material with this article. The
per-value provenance of
tables~\ref{tab:models}, \ref{tab:hardware} and~\ref{tab:devices},
recording the page of the source article each entry was read from, is
supplied alongside them, together with the record of the bibliography
verification.}

\bibliographystyle{unsrt}
\bibliography{refs}

\end{document}